\documentclass[12pt,english]{elsarticle}
\usepackage[LGR,T1]{fontenc}
\usepackage[latin9]{inputenc}
\usepackage{geometry}
\usepackage{textcomp}
\usepackage{amstext}
\usepackage{graphicx}

\makeatletter
\usepackage{babel}

\usepackage{babel}

\makeatother

\usepackage{babel}
\begin{document}
\begin{frontmatter}

\title{Structures and transitions in bcc tungsten grain boundaries and their
role in the absorption of point defects}

\author{Timofey Frolov$^{1}$, Qiang Zhu$^{2}$, Tomas Oppelstrup$^{1}$,
Jaime Marian$^{3}$ and Robert E. Rudd$^{1}$}

\address{$^{1}$ Lawrence Livermore National Laboratory, Livermore, California
94550, USA}

\address{$^{2}$Department of Physics and Astronomy, High Pressure Science
and Engineering Center, University of Nevada, Las Vegas, Nevada 89154,
USA}

\address{$^{3}$ Department of Materials Science and Engineering, University
of California Los Angeles, Los Angeles, California 90095, USA}
\begin{abstract}
We use atomistic simulations to investigate grain boundary (GB) phase
transitions in elemental body-centered cubic (bcc) metal tungsten.
Motivated by recent modeling study of grain boundary phase transitions
in {[}100{]} symmetric tilt boundaries in face-centered cubic (fcc)
copper, we perform a systematic investigation of {[}100{]} and {[}110{]}
symmetric tilt high-angle and low-angle boundaries in bcc tungsten.
The structures of these boundaries have been investigated previously
by atomistic simulations in several different bcc metals including
tungsten using the the $\gamma$-surface method, which has limitations.
In this work we use a recently developed computational tool based
on the USPEX structure prediction code to perform an evolutionary
grand canonical search of GB structure at 0~K. For high-angle {[}100{]}
tilt boundaries the ground states generated by the evolutionary algorithm
agree with the predictions of the $\gamma$-surface method. For the
{[}110{]} tilt boundaries, the search predicts novel high-density
low-energy grain boundary structures and multiple grain boundary phases
within the entire misorientation range. Molecular dynamics simulation
demonstrate that the new structures are more stable at high temperature.
We observe first-order grain boundary phase transitions and investigate
how the structural multiplicity affects the mechanisms of the point
defect absorption. Specifically, we demonstrate a two-step nucleation
process, when initially the point defects are absorbed through a formation
of a metastable GB structure with higher density, followed by a transformation
of this structure into a GB interstitial loop or a different GB phase. 
\end{abstract}
\end{frontmatter}

\section{Introduction}

Grain boundaries (GBs) greatly influence many properties of engineering
materials~\citep{Balluffi95}. Materials with high volume fraction
of GBs such as nano-crystalline and ultra-fine grain materials promise
improved strength~\citep{WEI20064079,Chookajorn951} and higher radiation
tolerance~\citep{El-Atwani:2017aa,ElAtwani201568,Bai26032010,Uberuaga:2015aa}.
As such they are potential candidates for materials that can operate
in extreme conditions. Many energy related applications place unique
demands on materials. For example in fusion, first-wall materials
must withstand the thermal load and the radiation field while maintaining
structural integrity both in terms of mechanical properties and in
terms of resisting erosion into the plasma due to plasma-materials
interaction. Tungsten has been identified as the divertor material
in ITER~\citep{doi:10.1146/annurev-matsci-070813-113627} and is
a leading candidate for the plasma-facing components in DEMO~\citep{0029-5515-47-11-014}
and subsequent magnetic fusion energy systems. It has a number of
advantageous properties: high thermal conductivity, acceptable activation
levels, high melting temperature, mechanical strength at elevated
temperatures, and resistance to surface sputtering. The questions
of recrystallization and embrittlement are particularly important~\citep{Mutoh95}.
Plasma facing components must operate below their recrystallization
temperature and they need to be replaced before undergoing brittle
failure. Tungsten is known to be susceptible to embrittlement. Below
its ductile-brittle transition temperature (\textasciitilde{}500 $^{\circ}$C),
pristine tungsten undergoes fracture by cleavage with essentially
no plasticity prior to failure. At higher temperatures, recrystallization
takes place, and GB embrittlement is the dominant fracture mode. Radiation
damage affects tungsten's failure properties. Predictive modeling
of recrystallization and deformation of polycrystalline W relies on
the accurate description of the W GBs. The goal of this work is to
use atomistic modeling to investigate the structure of bcc W GBs and
potential effects of point defects and elevated temperatures.

A growing number of recent studies suggests that GBs can exist in
multiple different states or phases and exhibit first-order structural
transformations in which the properties such as mobility, sliding
resistance and solute segregation change discontinuously~\citep{Cantwell20141}.
Experiments have revealed a potentially important role of GB phase
transitions~\citep{Dillon20076208,Cantwell20141} in abnormal grain
growth in ceramics~\citep{Dillon20076208}, activated sintering \citep{Luo99}
and liquid metal embrittlement \citep{Luo23092011}. Experimental
investigation of the potential impact of GB phase transitions on microstructure
and other materials properties is currently an active area of research
\citep{Baram08042011,Luo23092011,Harmer08042011,Rheinheimer201568,Cantwell20141,Pan201791,SCHULER2017196}.

Theoretically, GB phase transitions were investigated using phase-field
models that predicted a variety of possible transformations \citep{Tang06,Tang06b,Dillon20076208,ABDELJAWAD2017528,Mishin08a}.
A thermodynamic framework describing GB phase equilibrium and an adsorption
equation for GB phase junctions has been recently proposed \citep{Frolov:2015ab}.
Layering transitions associated with GB segregation were investigated
using lattice gas models \citep{Rickman201388,Rickman20161,Rickman2016225}
and first-principles calculations \citep{PhysRevB.90.144102}. Although
fundamentally important, the thermodynamic analysis \citep{Frolov:2015ab,Cahn82a,Rottman1988a,Hart:1972aa,Mishin08a}
and phase-field models \citep{Tang06,Tang06b,Dillon20076208,Mishin08a}
do not provide atomic-level details about the structures of different
GB phases and the mechanisms of first-order GB phase transitions.

Atomistic computer simulations have proven to be an invaluable tool
for the study of GBs. Such simulations have been applied to predict
GB structures and calculate their thermodynamic and kinetic properties,
such as GB free energies, diffusivities and mobilities as functions
of temperature and chemical composition \citep{Balluffi95,Ratanaphan2015346,Janssens06,Olmsted07a,Olmsted2011}.
In the common modeling approach, also known as the $\gamma$-surface
method, a GB is constructed by joining two perfect half-crystals with
different orientations while sampling the possible translations of
the grains relative to each other. This methodology has been employed
to predict structures and energies of GBs including those in bcc materials
\citep{KIM20111152,MORITA19971053,PhysRevB.85.064108,doi:10.1080/13642818908211183,doi:10.1080/01418619008244790,doi:10.1063/1.347741,PhysRevB.64.174101,HAHN2016108}.
The $\gamma$-surface approach has been challenged by a number of
computational studies of GBs in several different materials systems.
The studies demonstrated that the constant number of atoms in the
simulation cell and the periodic boundary conditions prohibit the
boundary from sampling all possible configurations and atoms have
to be added or removed from the GB core to achieve the lowest energy
configurations. These limitations became apparent in early studies
of GBs in ionic crystals \citep{doi:10.1080/01418618308243118}. For
example, in simulations of {[}001{]} twist boundaries in rock-salt-structured
oxides the conventional methodology generates GBs with ions of the
same charge overlapping at the GB plane. Strong Coulomb repulsion
between the ions makes these boundaries nearly unstable with respect
to dissociation into free surfaces \citep{Wolf82}. This prediction
was in apparent contradiction with experimental observations \citep{doi:10.1080/01418618208236206}.
Tasker and Duffy demonstrated that the energy of these twist boundaries
in oxides can be reduced significantly if a fraction of ions is removed
from the GB core \citep{doi:10.1080/01418618308243118,doi:10.1080/01418618608242811,DUFFY84a}.
They proposed several low-energy structures in which the ionic density
at the boundaries was optimized manually.

In face-centered cubic (fcc) metallic systems, simulations of GBs
in the grand canonical ensemble demonstrated changes in GB atomic
density and predicted GB structures with lower energy \citep{Phillpot1992,Phillpot1994}.
Ordered ground states of Si twist GBs were found by optimizing the
atomic density and sampling the GB structure with simulated annealing
\citep{Alfthan06,Alfthan07}. Genetic algorithms designed to explore
a diverse population of possible structures were applied to search
for low-energy structures in symmetric tilt Si GBs \citep{PhysRevB.80.174102}
and multicomponent ceramic GBs \citep{Chua:2010uq}.

In fcc metals new ground states and structural phase transformations
were found in GBs by performing high-temperature simulations with
the boundaries connected to a source/sink of atoms. Multiple GB phases
characterized by different atomic densities were found in high-angle
$\Sigma5(210)[001]$ and $\Sigma5(310)[001]$ GBs in Cu, Ag, Au and
Ni \citep{Frolov2013}. Specifically, the calculations predicted a
new GB phase called Split Kites, which has high atomic density and
complex structure with a periodic unit several times larger than that
of the conventional Kite phase. The new modeling methodology demonstrated
fully reversible transitions with varying the temperature and/or concentration
of impurities or point defects \citep{Frolov2013,Frolov2013PRL,PhysRevB.92.020103,Frolov2016}.
Both vacancies and interstitials were loaded into the GB in separate
simulations and triggered transitions between the grain boundary phases
with different atomic densities. This multiplicity of GB phases and
GB phase transitions was demonstrated for {[}001{]} symmetric tilt
GBs spanning the entire misorientation range in the same model of
Cu \citep{Zhu2018}. Continuous vacancy loading into general GBs in
Cu revealed lower energy states with different atomic density \citep{Demkowicz2015}.

In bcc metals, atomically ordered GB structures with high atomic density
were observed upon cyclic loading of interstitials into the $\Sigma5(210)[001]$
boundary in Mo \citep{Novoselov2016276}. However, the energies of
these states were much higher than the energy of the ground state,
making them unlikely candidates for stable GB phases. Statistical
properties and multiplicity of states have been investigated in a
large number of boundaries in Al, Si and W and also demonstrated the
importance of the grand canonical searches \citep{Han2017}. Specifically,
the study demonstrated that the energy of a $\Sigma5$ twist boundary
in W decreased upon varying the atomic density. New ground states
and grain boundary phase transformations have been demonstrated in
the $\Sigma27(552)[011]$ symmetric tilt and two $\Sigma5(001)$ twist
GBs in tungsten, tantalum and molybdenum \citep{Frolov2018Nanoscale}
using the evolutionary structure prediction method \citep{Zhu2018}. 

Motivated by these studies and the observation of GB phase transitions
in {[}001{]} symmetric tilt boundaries in Cu \citep{Frolov2013,Frolov2016,Zhu2018}
and the $\Sigma27(552)[011]$ symmetric tilt GB in tungsten \citep{Frolov2018Nanoscale},
in this work we conduct a systematic study of {[}001{]} and {[}011{]}
symmetric tilt boundaries in bcc tungsten. We construct the boundaries
at 0~K using a recently developed evolutionary grand canonical search
(EGCS) method \citep{Zhu2018} which is based on the USPEX code \citep{doi:10.1063/1.2210932}
and compare the results to the predictions of the $\gamma$-surface
approach. We also perform molecular dynamics simulations to investigate
the effects of high temperature and point defects on the GB structure
in the context of possible first-order GB phase transitions. For completeness,
the previously published structure calculations for the $\Sigma27(552)[011]$
GB will be presented together with the new results. The rest of the
paper is organized as follows. We describe the methodology of 0~K
GB structure calculations as well as the methodology of the high-temperature
molecular dynamics simulations in Section \ref{sec:Methodology-of-atomistic}.
We present the results of the simulations in Section \ref{sec:Results}.
Our findings are summarized and discussed in Section \ref{sec:Discussion}.

\section{Methodology of atomistic simulations\label{sec:Methodology-of-atomistic}}

\subsection{Model systems}

We have modeled tungsten GBs using two different embedded-atom method
(EAM) potentials: EAM1 \citep{Marinica2013} and EAM2 \citep{Zhou2001}.
While several W potentials are available in the literature, we selected
these because they gave better agreement with the existing DFT calculations
of GB energies \citep{Kurtz2014,swissW2016,Setyawan2012558}.

GB structure and energy calculations were performed for two different
sets of boundaries. The first set contained eighteen {[}001{]} symmetric
tilt boundaries with the misorientation angle $\theta$ ranging from
0 to $\pi/2$ radians. The second set contained fifty seven {[}110{]}
symmetric tilt boundaries, with the misorientation angle ranging from
$0$ to $\pi$. The boundaries were obtained by rotating the upper
and the lower grains around the common tilt axis by the angles $\theta/2$
and $-\theta/2$, respectively. The orientations of the reference
crystals were ($[100]$, $[010]$, $[001]$ ) and ($[110]$, $[001]$,
$[1\overline{1}0]$) for the {[}001{]} and the {[}110{]} sets of boundaries,
respectively. The boundaries were chosen to minimize the GB area for
computational efficiency, while evenly sampling the entire misorientation
angle range. The boundary normal was parallel to the $y$ direction
and the tilt axis was parallel to the $z$ direction of the simulation
block. Periodic boundary conditions are applied in the direction parallel
to the boundary. Periodic boundary conditions were not applied in
the direction normal to the boundary plane, so that the two bulk crystals
were terminated by two surfaces. GB structures and energies were calculated
at 0~K using the $\gamma$-surface approach as well as by the evolutionary
search \citep{Zhu2018}.

\subsection{$\gamma$-surface method}

In the $\gamma$-surface approach two perfect half-crystals with different
orientations are shifted relative to each other by a certain translation
vector and then joined together. The translation is followed by a
local relaxation of atoms that minimizes the energy of the system.
This procedure often yields several different metastable GB states
that correspond to different translation vectors. The configuration
with the lowest GB energy is assumed to be the ground state. The $\gamma$-surface
approach is relatively computationally inexpensive and often predicts
ground state structures \citep{CAMPBELL19993977}. However, it is
known to suffer from significant limitations. First, the search is
not grand canonical, which in this context means that no atoms are
inserted or removed from the GB core. Second, it does very poor sampling
of possible GB structures: during the energy minimization the atoms
simply fall into the local energy minima from their ideal lattice
positions and do not explore other configurations.

\subsection{Evolutionary Grand-Canonical Search (EGCS)}

In the second approach we constructed the GBs using a recently developed
evolutionary algorithm \citep{Zhu2018} based on the USPEX crystal
structure prediction code \citep{doi:10.1063/1.2210932}. USPEX has
proved to be extremely powerful in different systems including bulk
crystals \citep{doi:10.1063/1.2210932}, 2D crystals \citep{Zhou-PRL-2014},
surfaces \citep{Zhu-PRB-2013}, polymers \citep{Zhu-JCP-2014} and
clusters \citep{Lyakhov-CPC-2013}. The GB structure search algorithm
samples a wide range of different atomic structures, varies the GB
atomic density by inserting and removing atoms and explores different
GB dimensions to search for large-area reconstructions. In our implementation,
we split the bicrystal into three different regions, the upper grain
(UG), the lower grain (LG), and the GB region. We create the first
generation of GB structures by randomly populating GB regions with
atoms, imposing random layer group symmetries in different bicrystals,
and then joining the three regions together applying random relative
translations parallel to the GB plane. In the population different
bicrystals have different GB dimensions generated as random multiples
of the smallest periodic GB unit. The structures are then relaxed
externally by the LAMMPS code \citep{Plimpton95} and the GB energy
which serves as a fitness parameter is evaluated. During the optimization,
the atoms in the GB region are relaxed downhill fully, while the atoms
in the bulk only move as rigid bodies.

This population of different GB structures evolves over up to 50 generations.
Each new generation is produced from the previous one by operations
of heredity and mutation. Structures with low GB energy are more likely
to be selected as parents to produce the new child structures. In
the heredity operation two GB structures are randomly sliced and the
parts from different parents are combined to generate the offspring.
In a mutation operation the GB atoms displace according to the stochastically
picked soft vibrational modes based a bond-hardness model \citep{Zhu-PRB-2015,Lyakhov-CPC-2013}.
To sample different atomic densities atoms in the GB region are inserted
and deleted \citep{Lyakhov-CPC-2013,Zhu2018}. GB structures with
different dimensions are sampled automatically during the search by
replicating the existing bicrystals \citep{Zhu-PRB-2015}. The offspring,
together with a few best structures from the previous generation,
comprise the new population. This whole cycle is repeated until no
lower-energy structures are produced for sufficiently many generations.
A more detailed description of the algorithm can be found in Ref.~\citep{Zhu2018}.

The evolutionary search calculations are more computationally demanding
compared to the simple $\gamma$-surface approach. As a result, we
investigated only a subset of representative boundaries. Motivated
by the observations of GB phase transitions in Cu, out of the {[}001{]}
set we selected $\Sigma5(310)[001]$ and $\Sigma5(210)[001]$ boundaries,
which are the typical high-angle high-energy boundaries with misorientation
angles $\theta=36.87^{\circ}$ and $\theta=53.13^{\circ}$, respectively.
We also selected six {[}110{]} symmetric tilt boundaries: $\Sigma33(118)[1\overline{1}0]$
($\theta=20.1^{\circ}$), $\Sigma19(116)[1\overline{1}0]$ ($\theta=26.5^{\circ}$),
$\Sigma3(112)[1\overline{1}0]$ ($\theta=70.5^{\circ}$), $\Sigma3(111)[1\overline{1}0]$
($\theta=109.5^{\circ}$), $\Sigma3(332)[1\overline{1}0]$ ($\theta=129.5^{\circ}$)
and $\Sigma27(552)[1\overline{1}0]$ ($\theta=148.4^{\circ}$). These
boundaries sample the entire misorientation range $0<\theta<\pi$
and have been investigated recently by DFT calculations \citep{Kurtz2014,swissW2016,Setyawan2012558}.

\subsubsection{High-temperature simulations}

To validate the ground state structures predicted at 0~K, we performed
molecular dynamics simulations at high temperatures with GBs terminated
at open surfaces following the methodology introduced in Ref.~\citep{Frolov2013}.
Open surfaces provide a source and sink for atoms and effectively
introduces grand canonical environment in the GB region. The simulations
were performed in the temperature range from 1000~K to 3000~K. Typical
dimensions of the simulation blocks were $25.0\times20\times6$ nm$^{3}$.
In the $x$ direction the bicrystals were terminated by two open surfaces.
Periodic boundary conditions were applied only along the $z$ direction
which is parallel to the tilt axis. In the direction normal the boundary
plane the simulation block was terminated by two boundary regions
that were kept fixed during the simulation. We used the GB structures
generated by both the $\gamma$-surface method and the evolutionary
search as the initial configurations for the molecular dynamics simulations
to ensure that the final GB state is independent of the initial conditions.
The simulations were performed in the NVT canonical ensemble with
Nose-Hoover thermostat for up to 200 ns.

To investigate how changes in the GB atomic density affect GB structure
at finite temperature and demonstrate the mechanisms of point defect-GB
interaction, we performed isothermal simulations with the $\Sigma5(310)[001]$
and the $\Sigma27(552)[0\overline{1}1]$ GBs using periodic boundary
conditions along the boundary plane. In these simulations the interstitial
atoms were injected in the bulk crystal 5 to 10~{\AA } above the
GB plane. The simulations were performed at temperatures of 2000~K
and 2500~K for several tens of nanoseconds.

In the case of the $\Sigma27(552)[0\overline{1}1]$ boundary we simulated
coexistence of two different GB phases in a closed system at 1500
K, 1800 K, 2000 K and 2500 K for up to 200 ns. For the coexistence
simulations we use a larger block with dimensions $49.5\times2.7\times13.0$
nm$^{3}$. The heterogeneous two GB state was obtained again by injecting
interstitials into a half of the simulation block.

\section{Results\label{sec:Results}}

\subsection{GB structures and energies from the $\gamma$-surface approach}

\subsubsection{{[}001{]} symmetric tilt boundaries}

\begin{figure}
\includegraphics[width=1\textwidth]{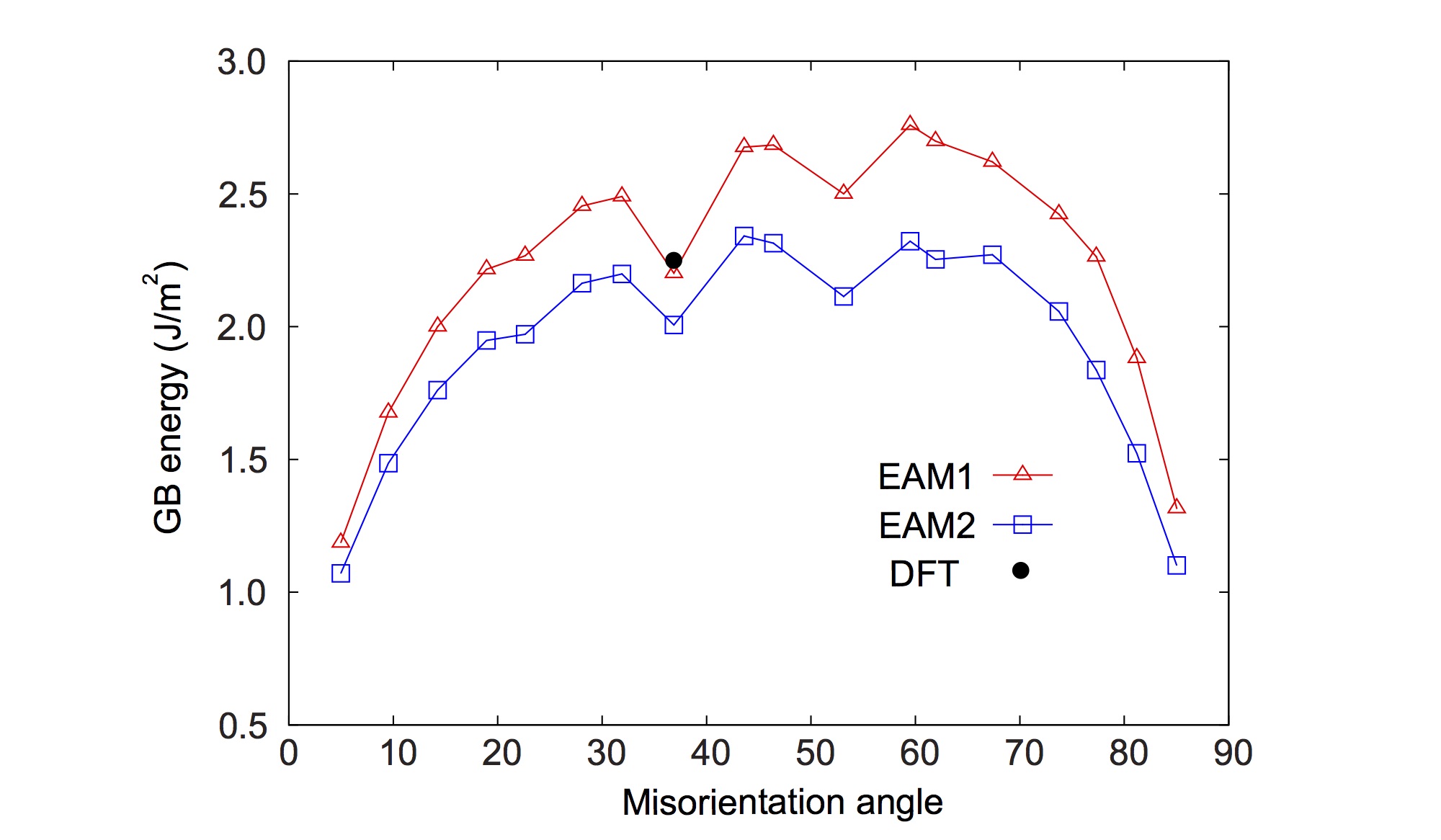}

\protect\protect\protect\caption{Energy of $\gamma$-surface generated boundaries: {[}100{]} symmetric
tilt boundaries. The plot shows the GB energy as a function of the
misorientation angle $\theta$ predicted by the EAM1 and EAM2 potentials.
The DFT calculated energy of the $\Sigma5(310)[001]$ boundary at
$\theta=36.87^{\circ}$ is taken from Ref.~\citep{swissW2016}. \label{fig:Egb_100tilt_gamma}}
\end{figure}

Fig.~\ref{fig:Egb_100tilt_gamma} illustrates GB energy of the {[}001{]}
symmetric tilt boundaries as a function of the misorientation angle
$\theta$ generated using the $\gamma$-surface approach with the
EAM1 and EAM2 potentials. The two energy cusps at $\theta=36.87^{\circ}$
and $\theta=53.13^{\circ}$ correspond to the $\Sigma5(310)[001]$
and $\Sigma5(210)[001]$ boundaries, respectively. The structures
of these boundaries, illustrated in Fig.~\ref{fig:GB-structures_frac0}(a
and b), are well known and are composed of kite-shaped structural
units. The left-hand side panel shows GB structure with the tilt axis
normal to plane of the figure, while in the right-hand side panel
the tilt axis is parallel to the plane of the figure. Both potentials
predict similar shape of the energy curve, but the magnitude of the
GB energy is somewhat different for the two potentials. The EAM1 potential
due to Marinica et al.~\citep{Marinica2013} shows an excellent agreement
with the DFT calculations of $\Sigma5(210)[001]$ boundary from Refs.~\citep{Kurtz2014,swissW2016,Setyawan2012558}.

\subsubsection{{[}110{]} symmetric tilt boundaries}

\begin{figure}
\begin{centering}
\includegraphics[width=1\textwidth]{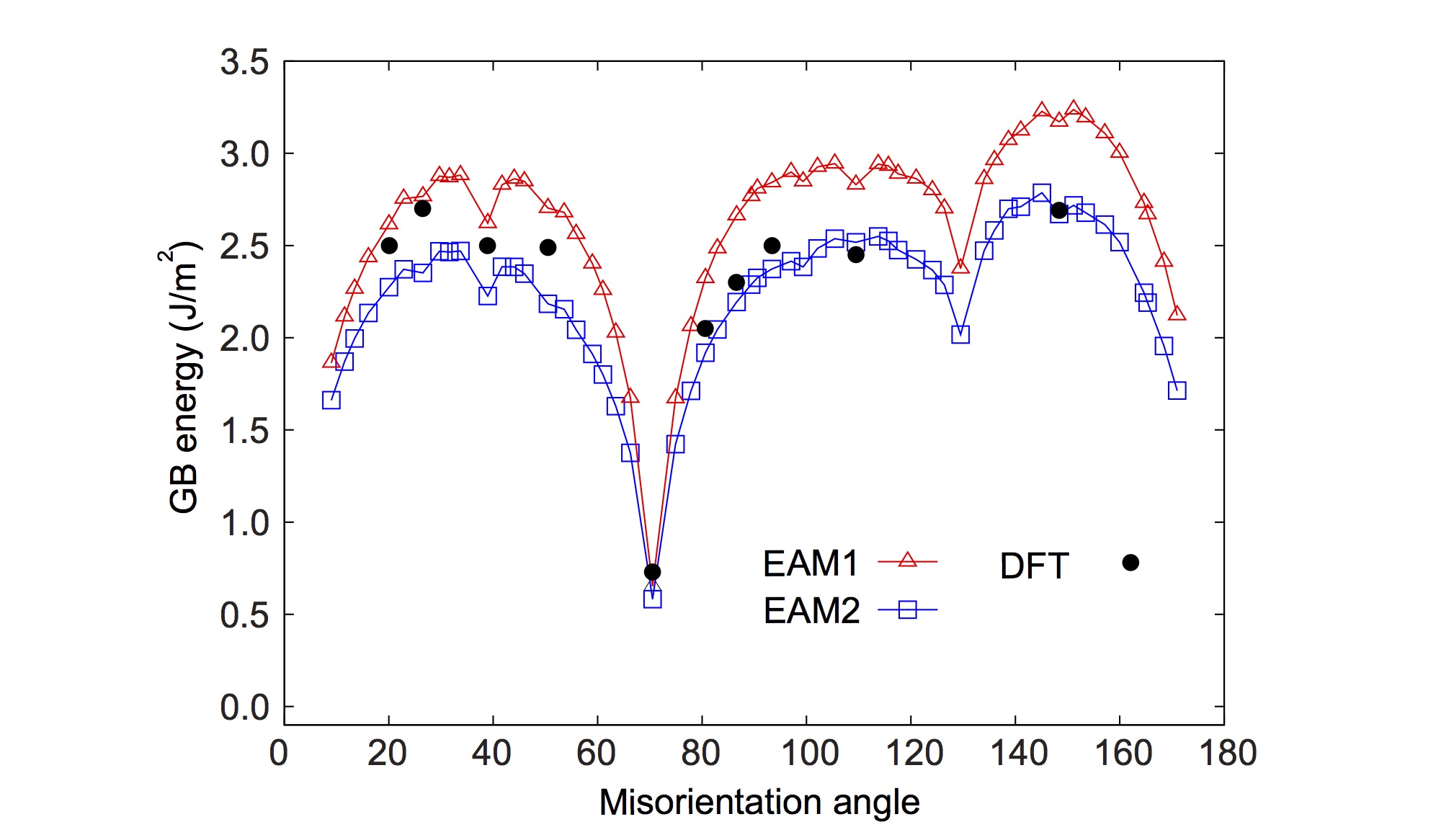} 
\par\end{centering}

\protect\protect\protect\caption{Energy of $\gamma$-surface generated boundaries: {[}110{]} symmetric
tilt boundaries. The plot shows the GB energy as a function of the
misorientation angle $\theta$ predicted by the EAM1 and EAM2 potentials.
The DFT calculated energies are taken from Ref.~\citep{swissW2016}.
\label{fig:Egb-110gamma}}
\end{figure}

Fig.~\ref{fig:Egb-110gamma} illustrates GB energy as a function
of the misorientation angle $\theta$ calculated for the {[}110{]}
symmetric tilt boundaries using the $\gamma$-surface approach with
the EAM1 and EAM2 potentials. The energies of a large set of {[}110{]}
symmetric tilt boundaries, generated using the same methodology, were
previously calculated for bcc W, Mo and Fe using DFT calculations
\citep{Kurtz2014,swissW2016}. The W data points from this study are
included in Fig.~\ref{fig:Egb-110gamma} for comparison. It is evident
from the figure that the two different potentials predict similar
trends in the GB energy as a function of the angle $\theta$, but
the magnitude of the energy is different. Both potentials agree reasonably
well with the DFT data \citep{Kurtz2014,swissW2016}. The GB energy
curve has two deep cusps at $\theta=70.5^{\circ}$ and $\theta=129.5^{\circ}$.
The deepest energy cusp at $\theta=70.5^{\circ}$ corresponds to the
$\Sigma3(112)[1\overline{1}0]$ boundary. The structure of this boundary
is illustrated in Fig.~\ref{fig:GB-structures_frac0}c.

\begin{figure}
\includegraphics[width=1\textwidth]{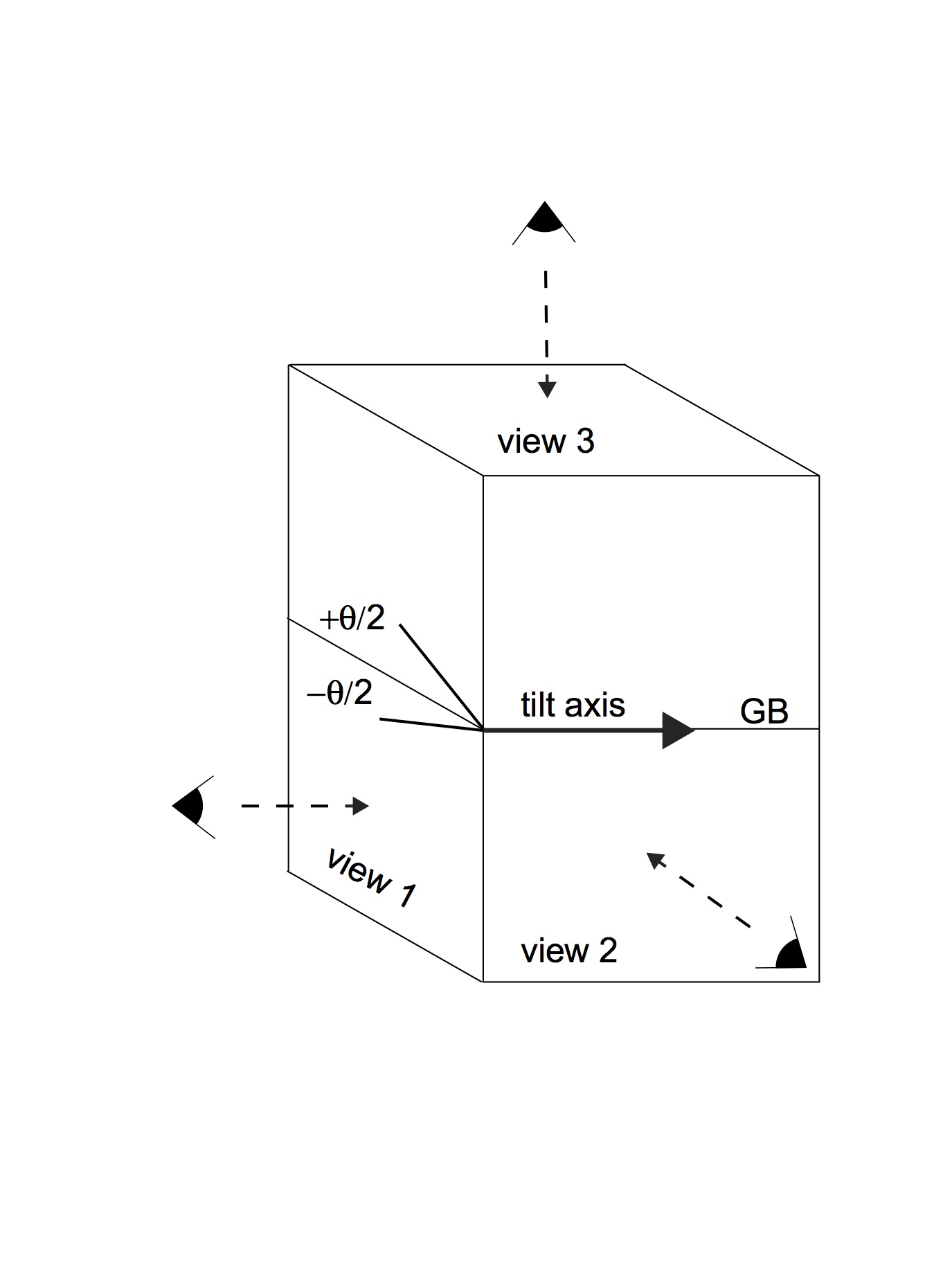}

\protect\protect\protect\caption{Schematic image of a bicrystal showing the upper and lower grains
misoriented by angles +$\theta$/2 and \textminus $\theta$/2 around
the common tilt axis and joined along the GB. The bicrystal and the
GB structure can be viewed from three different angles which provide
complementary information about the atomic structure. The three different
views are indicated on the image. GB structures generated in this
work are shown from these three views. \label{fig:GB-view-schem}}
\end{figure}

\begin{figure}
\includegraphics[width=1\textwidth]{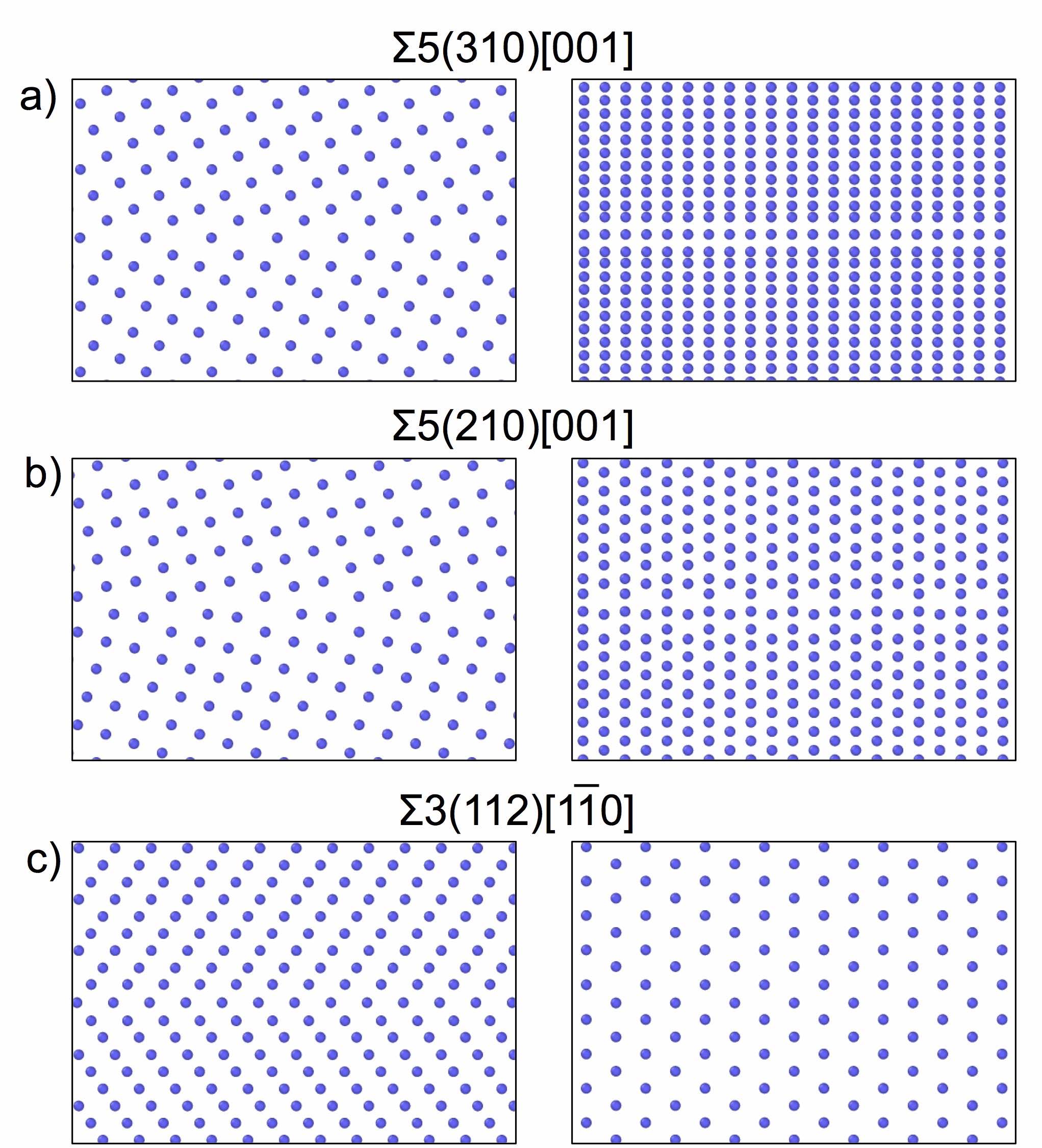}

\protect\protect\protect\caption{Ground state structures of the (a) $\Sigma5(310)[001]$, (b) $\Sigma5(210)[001]$
and (c) $\Sigma3(112)[1\overline{1}0]$ symmetric tilt boundaries
obtained by the $\gamma$-surface construction and the evolutionary
search. The left-hand panels correspond to view 1 as shown in Fig.~\ref{fig:GB-view-schem};
the right-hand panels correspond to view 2. The well-known structures
in a and b are composed of kite-shaped structural units. \label{fig:GB-structures_frac0}
In all three structures the GB atoms are confined to the abutting
(100) and (110) planes.}
\end{figure}

Despite the similarity in the functional form of the energy curves
predicted by the two potentials using the $\gamma$-surface approach,
in some cases, the potentials predicted very different GB structures
for the same misorientation angle. For example, Fig.~\ref{fig:GB-phases-20}(a
and b) illustrates two different structures of the $\Sigma33(118)[1\overline{1}0]$
$(20.1^{\circ})$ boundary predicted by EAM1 and EAM2 potentials,
respectively. The left-hand side, the middle and the right-hand side
panels of the figure show three different views of the GB structure.
The different views are explained in a schematic in Fig.~\ref{fig:GB-view-schem}.
In Fig.~\ref{fig:GB-phases-20} the structural units of both configurations
predicted by the two potentials are outlined by a red line to guide
the eye. The EAM1 structure agrees with the DFT calculations from
Ref.~\citep{swissW2016}. While empirical potentials are not perfect
and may predict different defect structures, below we demonstrate
that the discrepancy in the predicted structure of the $\Sigma33(118)[1\overline{1}0]$
GB is due to the limitations of the $\gamma$-surface method.

\subsection{Evolutionary search}

\begin{figure}
\begin{centering}
\includegraphics[width=0.9\textwidth]{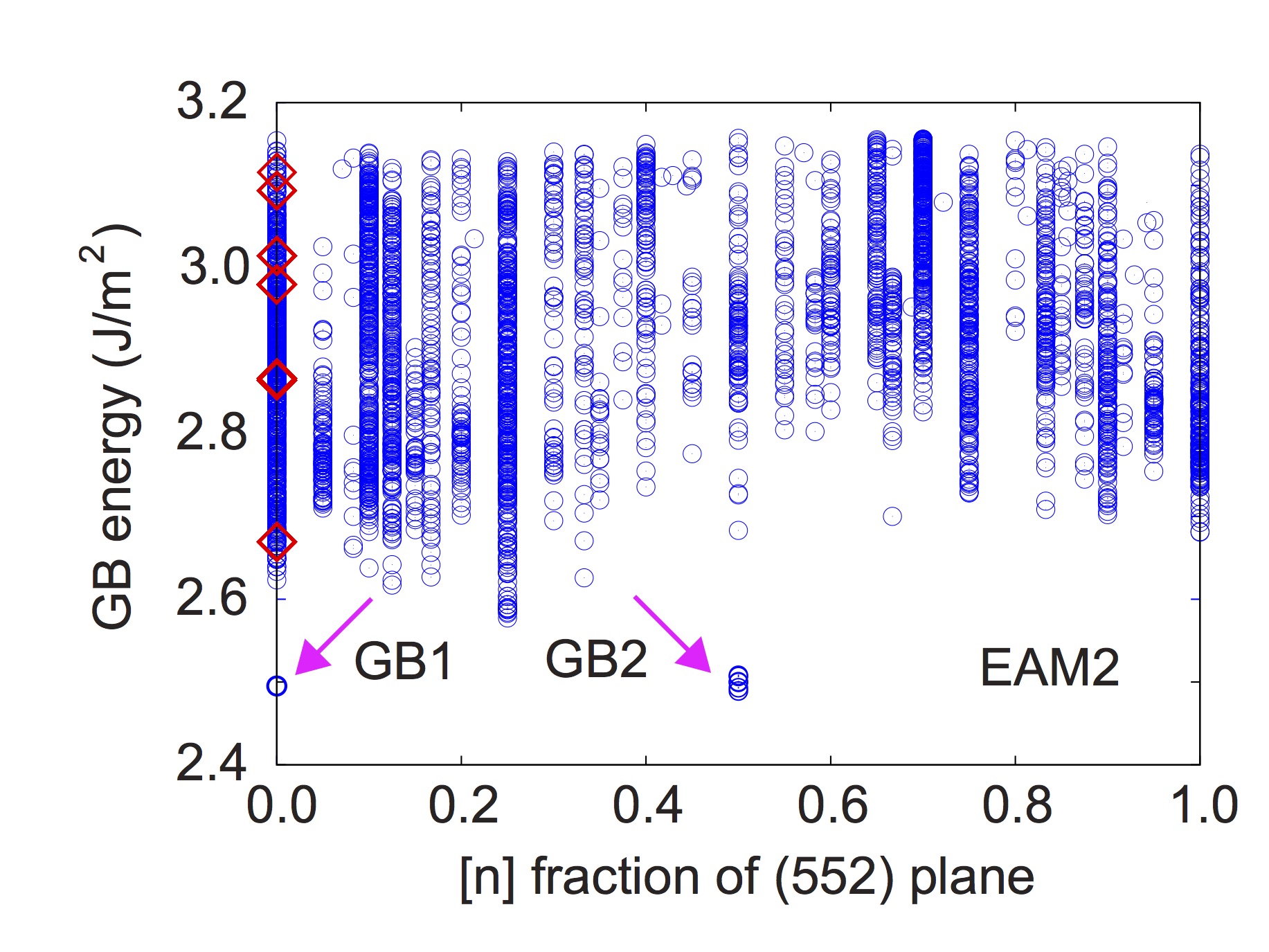} 
\par\end{centering}

\protect\protect\protect\caption{The evolutionary grand canonical search for the $\Sigma27(552)[1\overline{1}0]$
($\theta=148.1^{\circ}$) GB modeled with the EAM2 potential. GB energy
of different structures generated by the algorithm (blue circles)
is plotted as a function of the atomic density {[}n{]} measured as
a fraction of atoms in a (552) bulk atomic plane. The search finds
two low-energy GB phases at {[}n{]}=0 and {[}n{]}=0.5. Red diamonds
illustrate the GB structures generated using the conventional $\gamma$-surface
approach. Because no atoms are inserted or removed in this approach,
all red diamonds are located at $[n]=0$. Even without changing the
GB atomic density, the evolutionary search finds low-energy boundaries
missed by the conventional methodology. \label{fig:USPEX-vs-gamma-search}}
\end{figure}

We performed the evolutionary grand canonical structure search for
a subset of eight GBs which included two {[}001{]} tilt boundaries
and six {[}110{]} tilt boundaries. During the search the algorithm
explores different atomic densities of the GB core by inserting and
removing atoms. As a result, for each boundary the energy of different
structures can be plotted as a function of the number of inserted
atoms. Fig.~\ref{fig:USPEX-vs-gamma-search} illustrates the results
of the EGCS for the $\Sigma27(552)[1\overline{1}0]$ boundary modeled
with the EAM2 potential. Each blue circle on the plot represents a
GB structure generated by the evolutionary algorithm. The energy is
plotted as a function of number of atoms {[}n{]} measured as a fraction
of atoms in a (552) plane. To compare the results of the evolutionary
search with the predictions of the common methodology we included
the data points generated by the $\gamma$-surface method, which are
shown on the plot as red diamonds. Notice that all the red diamonds
are located at $[n]=0$ because the $\gamma$-surface method does
not insert or remove atoms from the GB core. The different energy
values correspond to the different rigid translations of the grains
relative to each other.

It is clear that the evolutionary search explores a much more diverse
space of GB configurations. For this particular boundary it finds
two distinct low-energy structures indicated by arrows at $[n]=0$
and $[n]=0.5$. At $[n]=0$ the evolutionary algorithm predicts the
lowest energy $\gamma_{\text{GB}}=$2.495 J/m$^{2}$, while the best
GB structure generated by the $\gamma$-surface method has a significantly
higher energy of $\gamma_{\text{GB}}=$2.67 J/m$^{2}$. In this case,
the 7\% reduction in energy is achieved by simply rearranging the
structure, because no atoms have been added or removed. This example
clearly demonstrates the insufficiency of the $\gamma$-surface method.
In addition to the rearrangement of the atoms, insertion and deletion
of atoms in the GB core enables the exploration of other potentially
important states such as a new ground state at {[}n{]}=0.5 with the
energy $\gamma_{\text{GB}}=$2.493 J/m$^{2}$. The low-energy structures
at {[}n{]}=0 and {[}n{]}=0.5 represent two different phases of the
$\Sigma27(552)[1\overline{1}0]$ GB.

\subsubsection{EGCS for {[}001{]} symmetric tilt boundaries}

Fig.~\ref{fig:USPEX-energy-100sh}(a and b) illustrates the results
of the evolutionary search performed for the $\Sigma5(210)[001]$
and $\Sigma5(310)[001]$ GBs, respectively. In both cases the lowest
energy configurations were found at {[}n{]}=0 and matched the ground
states generated by the conventional methodology. The energy of other
GB configurations increased with the increasing atomic density $[n]$
and reached the highest value at $[n]=0.5$ for the $\Sigma5(310)[001]$
boundary. These results suggest that the ground states composed of
kite-shaped structural units are stable against transformation to
structures with other densities. The well-known ground state structures
of these boundaries are illustrated in Fig.~\ref{fig:GB-structures_frac0}(a
and b). In the left-hand side panels the {[}001{]} tilt axis is normal
to the plane of the figure, while it is in the plane of the figure
on the right-hand side. In the schematic picture of the bicrystal
in Fig.~\ref{fig:GB-view-schem} this two views correspond to views
1 and 2, respectively. In both boundaries the atoms are confined to
(001) atomic planes of the abutting crystals.

\subsubsection{EGCS for {[}110{]} symmetric tilt boundaries}

\begin{figure}
\begin{centering}
\includegraphics[height=0.7\textheight]{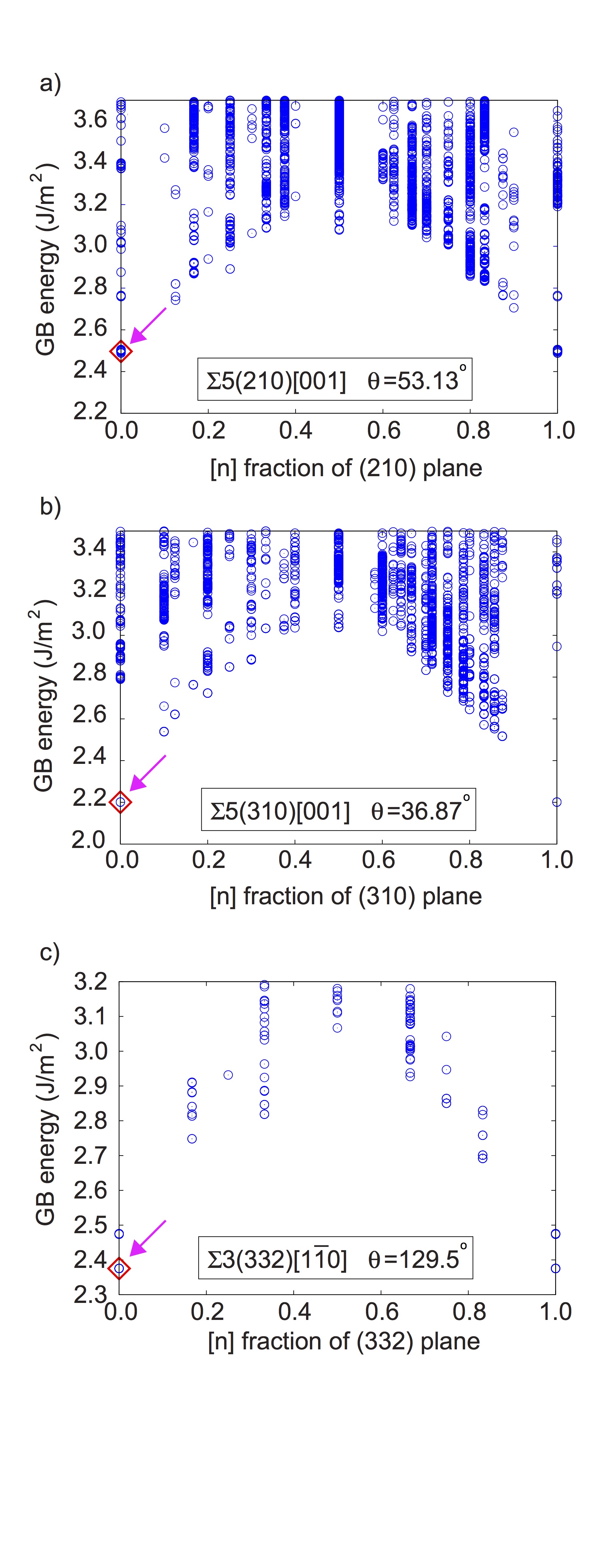} 
\par\end{centering}

\protect\protect\protect\caption{Results of the evolutionary grand canonical structure search for the
(a) $\Sigma5(210)[001]$, (b) $\Sigma5(310)[001]$ and (c) $\Sigma3(332)[1\overline{1}0]$
GBs modeled with the EAM1 potential. GB energy of different structures
generated by the algorithm (blue circles) is plotted as a function
of the atomic density {[}n{]} measured as a fraction of atoms bulk
atomic plane parallel to the boundary. In all three cases the ground
states were found at {[}n{]}=0. These ground states were also generated
by the conventional methodology (red diamonds). For these boundaries
the evolutionary search does not predict alternative low-energy phases
with different {[}n{]}. \label{fig:USPEX-energy-100sh}}
\end{figure}

\begin{figure}
\begin{centering}
\includegraphics[height=0.7\textheight]{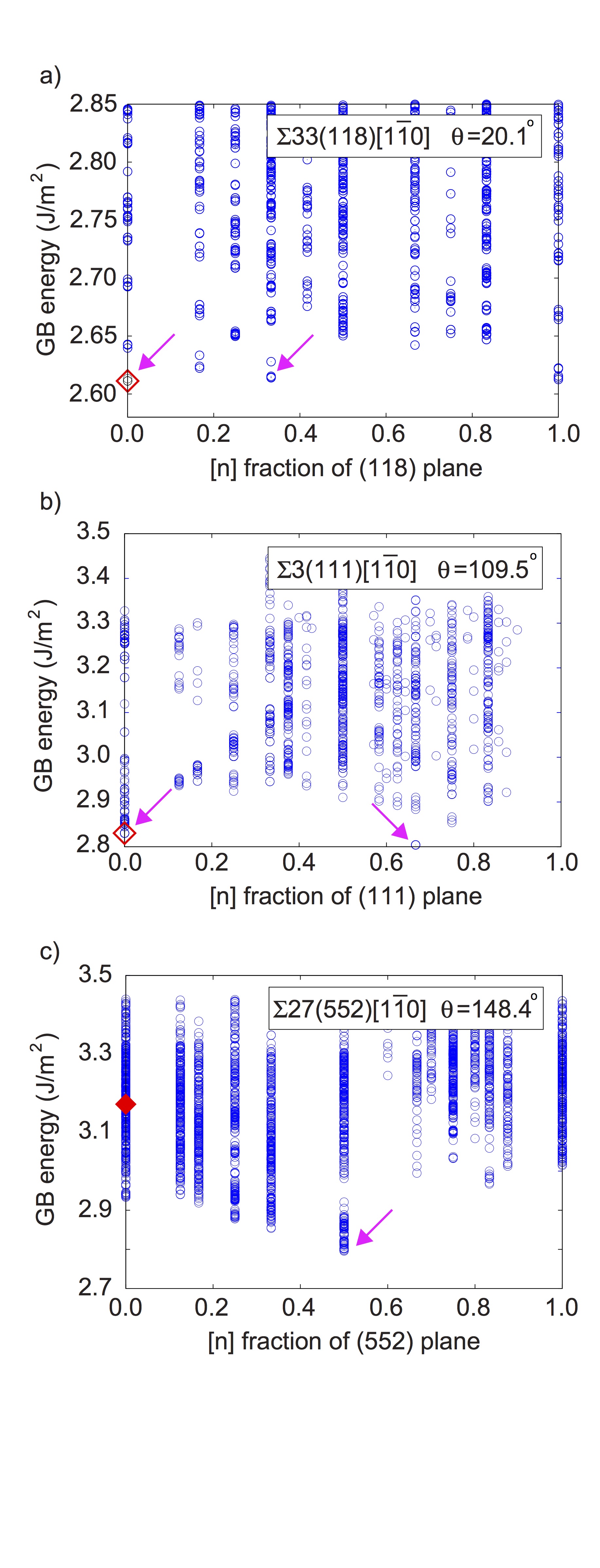} 
\par\end{centering}

\protect\protect\protect\caption{Results of the evolutionary grand canonical structure search for the
(a) $\Sigma33(118)[1\overline{1}0]$ ($\theta=20.1^{\circ}$), (b)
$\Sigma3(111)[1\overline{1}0]$ ($\theta=109.5^{\circ}$) and (c)
$\Sigma27(552)[1\overline{1}0]$ ($\theta=148.4^{\circ}$) GBs using
the EAM1 potential. These representative low-angle and high-angle
boundaries were selected to sample the entire misorientation range
$0^{\circ}<\theta<180^{\circ}$. GB energy of different structures
generated by the algorithm (blue circles) is plotted as a function
of the atomic density {[}n{]} measured as a fraction of atoms bulk
atomic plane parallel to the boundary. Red diamonds represent the
lowest energy states generated by the $\gamma$-surface approach.
For all three boundaries the evolutionary search predict alternative
low-energy structures with higher {[}n{]}. The different phases of
the boundaries have close energies and are indicated by magenta arrows.
\label{fig:USPEX-energy-110}}
\end{figure}

The evolutionary search conducted for four {[}110{]}-tilt boundaries
yielded additional GB structures that were significantly different
from those generated by the $\gamma$-surface approach. The studied
boundaries were selected from the entire misorientation range $0^{\circ}<\theta<180^{\circ}$
excluding the energy cusps located at $70.5^{\circ}$ and $129.5^{\circ}$.
Fig.~\ref{fig:USPEX-energy-110} illustrates the results of the evolutionary
structure search for GBs with $\theta=20.1^{\circ}$, $\theta=109.5^{\circ}$
and $\theta=148.1^{\circ}$ using the EAM1 potential. In contrast
to the searches shown in Fig.~\ref{fig:USPEX-energy-100sh}, each
of these boundaries exhibits a minimum at atomic densities other than
$[n]=0$, suggesting possible GB phases beyond those predicted by
the conventional methodology.

\begin{figure}
\begin{centering}
\includegraphics[width=0.8\textwidth]{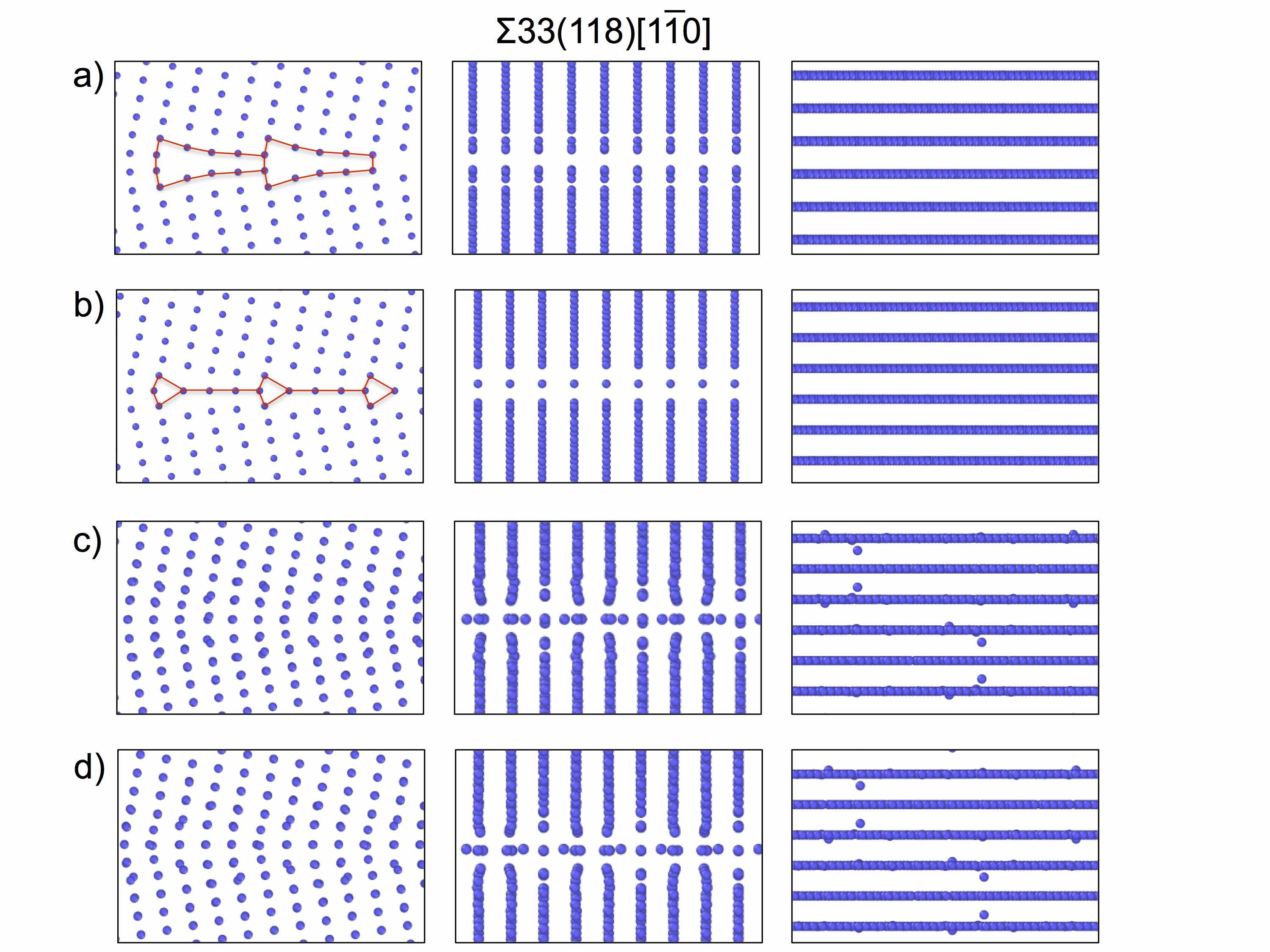} 
\par\end{centering}

\protect\protect\protect\caption{Structures of the $\Sigma33(118)[1\overline{1}0]$ ($\theta=20.1^{\circ}$)
GB generated by the conventional $\gamma$-surface approach and the
evolutionary grand canonical search. (a) The lowest energy state generated
using the $\gamma$-surface approach with the EAM1 potential with
energy $\gamma_{\text{GB}}=2.611$ J/m$^{2}$. This structure was
also predicted by prior DFT calculations~\citep{swissW2016,Kurtz2014}
that used the same methodology. (b) The lowest energy state generated
using the $\gamma$-surface approach with the EAM2 potential with
energy $\gamma_{\text{GB}}=2.257$ J/m$^{2}$. (c) GB structure predicted
by the evolutionary search with the EAM1 potential with the atomic
fraction {[}n{]}=1/3 and energy $\gamma_{GB}=2.615$ J/m$^{2}$. (d)
GB structure predicted by the evolutionary search using the EAM2 with
the same atomic fraction {[}n{]}=1/3 and energy $\gamma_{GB}=2.226$
J/m$^{2}$. The left-hand, middle and right-hand panels correspond
to views 1, 2 and 3 of the boundary, respectively. The different views
are described in Fig.~\ref{fig:GB-view-schem}. While the two potentials
predict different GB structures at {[}n{]}=0, new EGCS optimized states
at {[}n{]}=1/3 are the same. Views 2 and 3 reveal that in (c) and
(d) the GB atoms occupy positions in between (110) planes. \label{fig:GB-phases-20}}
\end{figure}

The GB energy cusps break the misorientation range into three intervals.
In the $0^{\circ}<\theta<70.5^{\circ}$ interval (Fig.~\ref{fig:Egb-110gamma}a)
we selected the $\theta=20.1^{\circ}$ and $\theta=26.5^{\circ}$
boundaries. These are relatively low-angle GBs composed of periodic
arrays of edge dislocations. Fig.~\ref{fig:USPEX-energy-110}a illustrates
the results of the evolutionary search for the $\Sigma33(118)[1\overline{1}0]$
boundary at $\theta=20.1^{\circ}$ modeled with the EAM1 potential.
The plot has two GB energy minima: one at $[n]=0$ and the second
one at {[}n{]}=1/3. The two low-energy configurations are indicated
by arrows on the plot. The search with the EAM2 potential predicted
similar behavior. At $[n]=0$ the evolutionary search yielded GB structures
identical to those generated by the $\gamma$-surface approach. As
discussed earlier, the EAM1 and EAM2 potentials predict different
ground states for the $\Sigma33(118)[1\overline{1}0]$ boundary, which
are illustrated in Fig.~\ref{fig:GB-phases-20}(a and b). The energies
of these states were $\gamma_{\text{GB}}=2.611$ J/m$^{2}$ and $\gamma_{\text{GB}}=2.257$
J/m$^{2}$ for the EAM1 and EAM2 potentials, respectively.

On the other hand, at $[n]=1/3$ with respect to the (118) plane,
the evolutionary search predicts a new GB structure with energies
$\gamma_{\text{GB}}=2.615$ J/m$^{2}$ and $\gamma_{\text{GB}}=2.226$
J/m$^{2}$ for EAM1 and EAM2 potentials, respectively. Thus, for each
potential the energies of the $[n]=1/3$ structure are nearly identical
to those of the $[n]=0$ structures. The $[n]=1/3$ EGCS structures
generated by EAM1 and EAM2 are illustrated in Fig.~\ref{fig:GB-phases-20}(c
and d). Remarkably, both potentials predict the same structure. The
$[n]=1/3$ configuration is a $1\times3$ reconstruction, which means
it has a larger unit cell compared to the $\gamma$-surface constructed
boundaries. The three different views of the GB structure reveal that
the extra atoms occupy interstitial positions within the GB plane.
This structural feature is very different from the conventional $[n]=0$
boundaries in which all atoms are confined to the (110) planes, as
can be seen in the middle and right-hand panels of Fig.~\ref{fig:GB-phases-20}(a
and b). Similar structures with higher atomic density {[}n{]} were
predicted by the evolutionary search for the $\Sigma19(116)[1\overline{1}0]$
GB at $\theta=26.5^{\circ}$.

\begin{figure}
\includegraphics[width=0.8\textwidth]{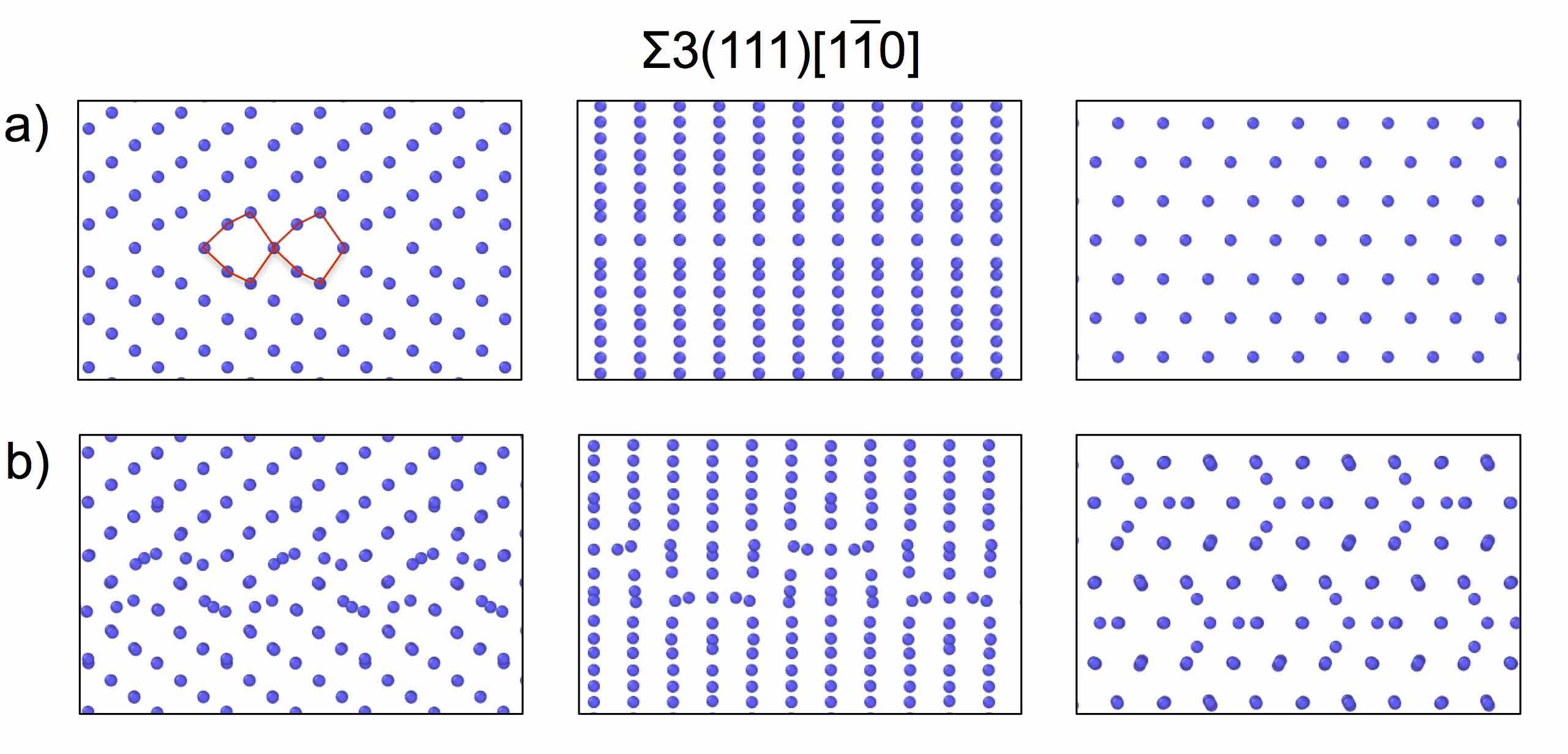}

\protect\protect\protect\caption{Two structures of the $\Sigma3(111)[1\overline{1}0]$ ($\theta=109.5^{\circ}$)
GB modeled with the EAM1 potential. (a) The structure predicted by
both the $\gamma$-surface approach and the evolutionary algorithm
at {[}n{]}=0 with energy $\gamma_{\text{GB}}=2.83$ J/m$^{2}$. (b)
The ground state predicted by the evolutionary algorithm at {[}n{]}=2/3
with energy $\gamma_{\text{GB}}=2.80$ J/m$^{2}$. The left-hand side,
the middle and the right-hand panels correspond to view 1, view 2
and view 3 of the boundary, respectively. The different views are
described in Fig.~\ref{fig:GB-view-schem}. Views 2 and 3 reveal
that the GB atoms occupy positions in between (110) planes. \label{fig:GB-phases-109}}
\end{figure}

Fig.~\ref{fig:USPEX-energy-110}b illustrates the results of the
evolutionary search with the EAM1 potential for the $\Sigma3(111)[1\overline{1}0]$
at $\theta=109.5^{\circ}$, which was selected as a representative
high-energy boundary from the $70.5^{\circ}<\theta<129.5^{\circ}$
interval. The energy plot again exhibits two distinct minima at $[n]=0$
and $[n]=2/3$ as indicated by the arrows on the plot, with the energies
$\gamma_{\text{GB}}=2.83$ J/m$^{2}$ and $\gamma_{\text{GB}}=2.80$
J/m$^{2}$, respectively. The {[}n{]}=2/3 is the ground state at 0~K,
but the energy difference between the two structures is only 1\%.
Fig.~\ref{fig:GB-phases-109}(a and b) shows the $[n]=0$ and $[n]=2/3$
structures of the boundary, respectively. The $[n]=0$ structure also
generated by the $\gamma$-surface approach can be described as composed
of kite-shaped structural units. The middle and right-hand panels
of Fig.~\ref{fig:GB-phases-109}a reveal that the atoms within the
boundary are confined to the misoriented $(1\overline{1}0)$ planes
of the two abutting grains. The {[}n{]}=2/3 phase is a $1\times3$
reconstruction, which means that the dimension of its smallest periodic
unit along the $[1\overline{1}0]$ tilt axis is three times larger
than that of the {[}n{]}=0 GB phase. The middle and the right-hand
panels of Fig.~\ref{fig:GB-phases-109}b demonstrate that the atoms
of the {[}n{]}=2/3 GB phase occupy sites between the misoriented $(1\overline{1}0)$
planes, forming an ordered structure within the GB plane.

\begin{figure}
\includegraphics[width=1\textwidth]{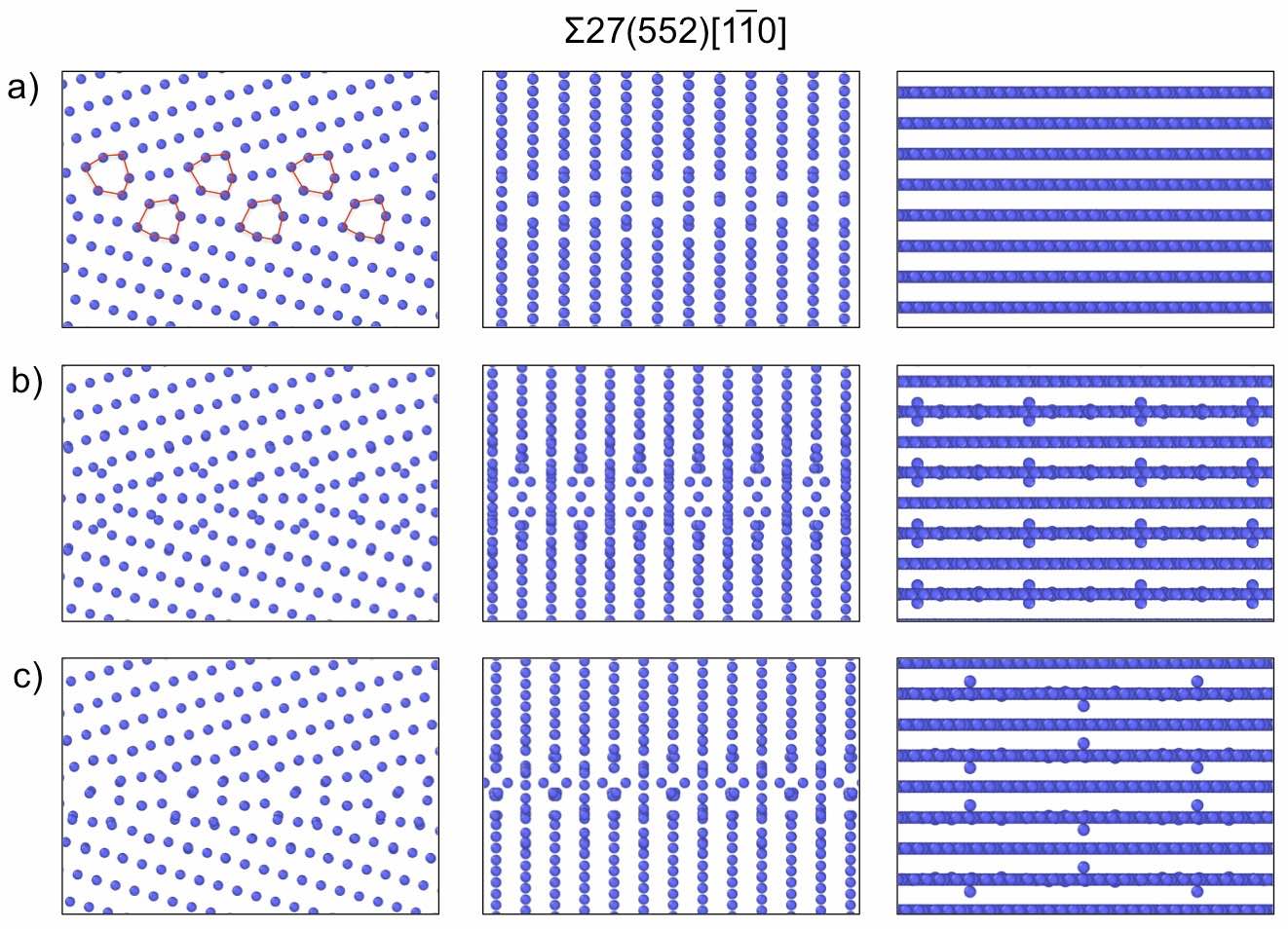}

\protect\protect\protect\caption{Multiple structures of the $\Sigma27(552)[1\overline{1}0]$ ($\theta=148.4^{\circ}$)
GB predicted by the EAM2 potential. (a) Best configuration predicted
by the conventional approach of $\gamma$-surface construction with
GB energy $\gamma_{GB}=2.67$ J/m$^{2}$. The evolutionary search
predicts GB phases {[}n{]}=0 (b) and {[}n{]}=1/2 (c) with energies
$\gamma_{GB}=2.495$ J/m$^{2}$ and $\gamma_{GB}=2.493$ J/m$^{2}$,
respectively. The left, middle and right panels show views 1, 2, and
3, respectively. Views 2 and 3 of (b) and (c) reveal the complex arrangement
of atoms within the boundary plane. \label{fig:GB-phases-148}}
\end{figure}

\begin{figure}
\includegraphics[width=1\textwidth]{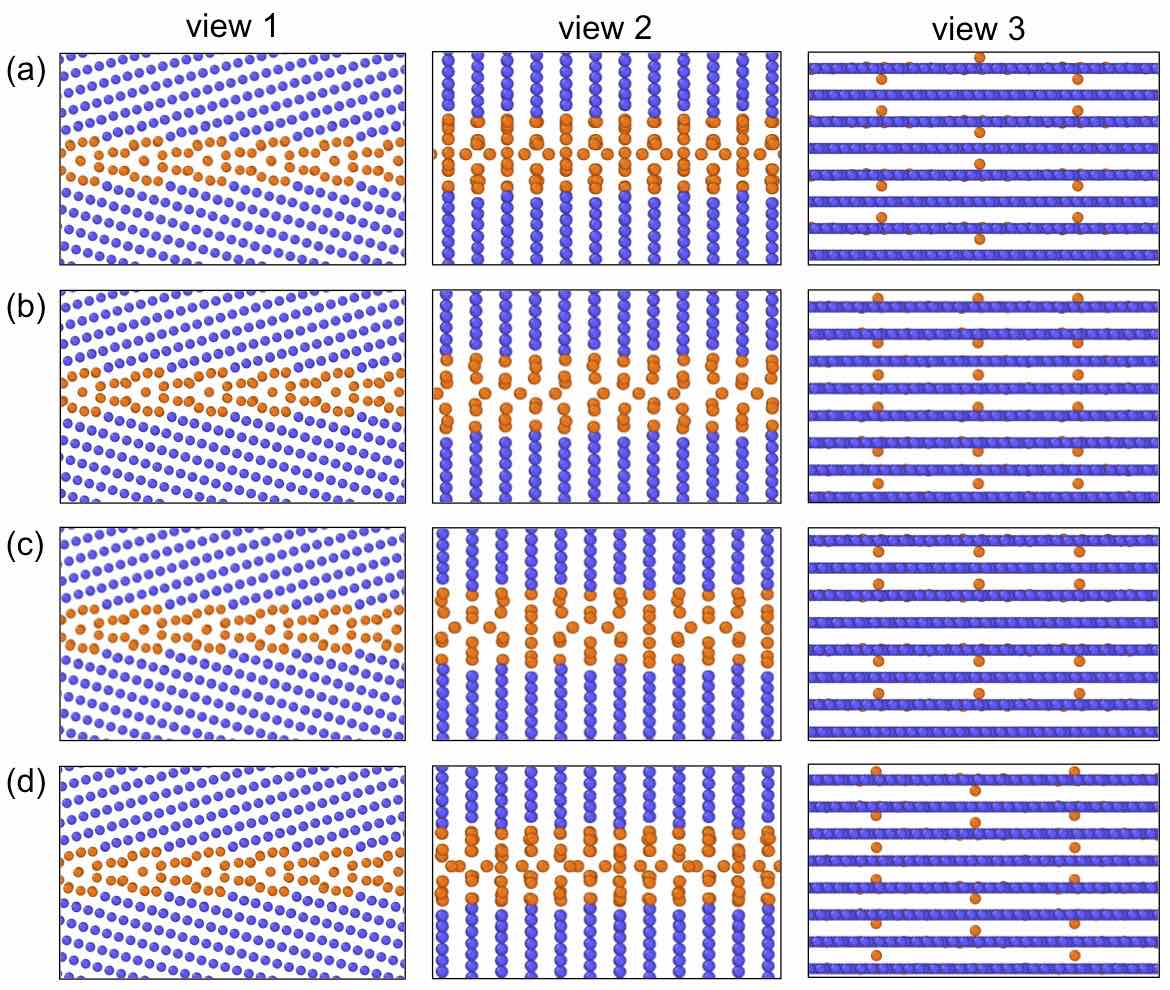}

\protect\protect\protect\caption{Multiple structures of the {[}n{]}=1/2 phase of the $\Sigma27(552)[1\overline{1}0]$
($\theta=148.4^{\circ}$) GB predicted by (a) EAM2 and (b-d) EAM1
potentials. The structures have nearly the same energy. The difference
in energy is less than the numerical accuracy of the calculations.
All structures are nearly indistinguishable when viewed in the left-hand
side panels (view 1). View 2 reveals that the structures are different.
View 3 showing the arrangement of the atoms within the GB plane reveals
that the different GB structures have different patterns formed by
the interstitial atoms. The GB atoms are identified according to common
neighbor analysis~\citep{0965-0393-18-1-015012}. \label{fig:GB-phases-148M_mult}}
\end{figure}

Finally, in the angle range $129.5^{\circ}<\theta<180^{\ensuremath{\circ}}$
we examined the $\Sigma27(552)[1\overline{1}0]$ boundary with at
$\theta=148^{\circ}$. Figs.~\ref{fig:USPEX-energy-110}c and \ref{fig:USPEX-vs-gamma-search}
illustrate the searches for this boundary modeled with the EAM1 and
EAM2 potentials, respectively. The predictions of the two potentials
are somewhat different. Specifically, the EAM2 predicts two distinct
low-energy GB phases located at $[n]=0$ and $[n]=0.5$, which were
discussed earlier and illustrated in Fig.~\ref{fig:GB-phases-148}(b
and c). On the other hand the EAM1 model predicts a single strong
minimum at $[n]=0.5$. The energy of this state, $\gamma_{\text{GB}}=2.81$
J/m$^{2}$, is 11\% lower than $\gamma_{\text{GB}}=3.17$ J/m$^{2}$
of the conventional structure generated by the $\gamma$-surface approach.
The $[n]=0.5$ structures are $1\times2$ reconstructions. The ground
states predicted by both potentials are not unique. Fig.~\ref{fig:GB-phases-148M_mult}(b-d)
illustrates several distinct structures of the $[n]=0.5$ GB phase
predicted using the EAM1 potential. The structure shown in Fig.~\ref{fig:GB-phases-148M_mult}a
was generated by the EAM2 potential. While the structures of these
boundaries look nearly indistinguishable in the left-hand side panels
of Fig.~\ref{fig:GB-phases-148M_mult}, the middle and the right-hand
panels clearly show different atomic arrangements. The main difference
between the structures is the pattern of the occupied interstitial
sites within the GB plane. Remarkably, all these configurations have
nearly the same energy. The difference lies within the numerical accuracy
of the calculations. The energy of these states was recently calculated
using DFT calculations which confirmed the predictions of the empirical
potentials EAM1 and EAM2~\citep{Zhu2018}.

The evolutionary search performed for the $\Sigma3(112)[1\overline{1}0]$
($\theta=70.5^{\circ}$) and $\Sigma3(332)[1\overline{1}0]$ ($\theta=129.5^{\circ}$)
GBs that correspond to the GB energy cusps in Fig.~\ref{fig:Egb-110gamma}
agreed with the $\gamma$-surface method and did not yield other alternative
low-energy configurations. An example of the evolutionary search for
the $\Sigma3(332)[1\overline{1}0]$ boundary modeled with the EAM1
potential is shown in Fig.~\ref{fig:USPEX-energy-100sh}c. It is
qualitatively similar to the searches for the {[}001{]}-tilt boundaries
with a single energy minimum located at the origin of the plot.

\subsection{Molecular dynamics simulations}

\subsubsection{High-temperature simulations with open surfaces}

Isothermal molecular dynamics simulations of the $\Sigma5(310)[001]$
tilt boundary with open surfaces confirmed that the structure calculated
at 0~K was also stable at high temperature. We conclude that, the
energy analysis at 0~K and the simulations at high temperature demonstrate
that the $\gamma$-surface approach accurately predicts the ground
state for this boundary.

\begin{figure}
\begin{centering}
\includegraphics[width=1\textwidth]{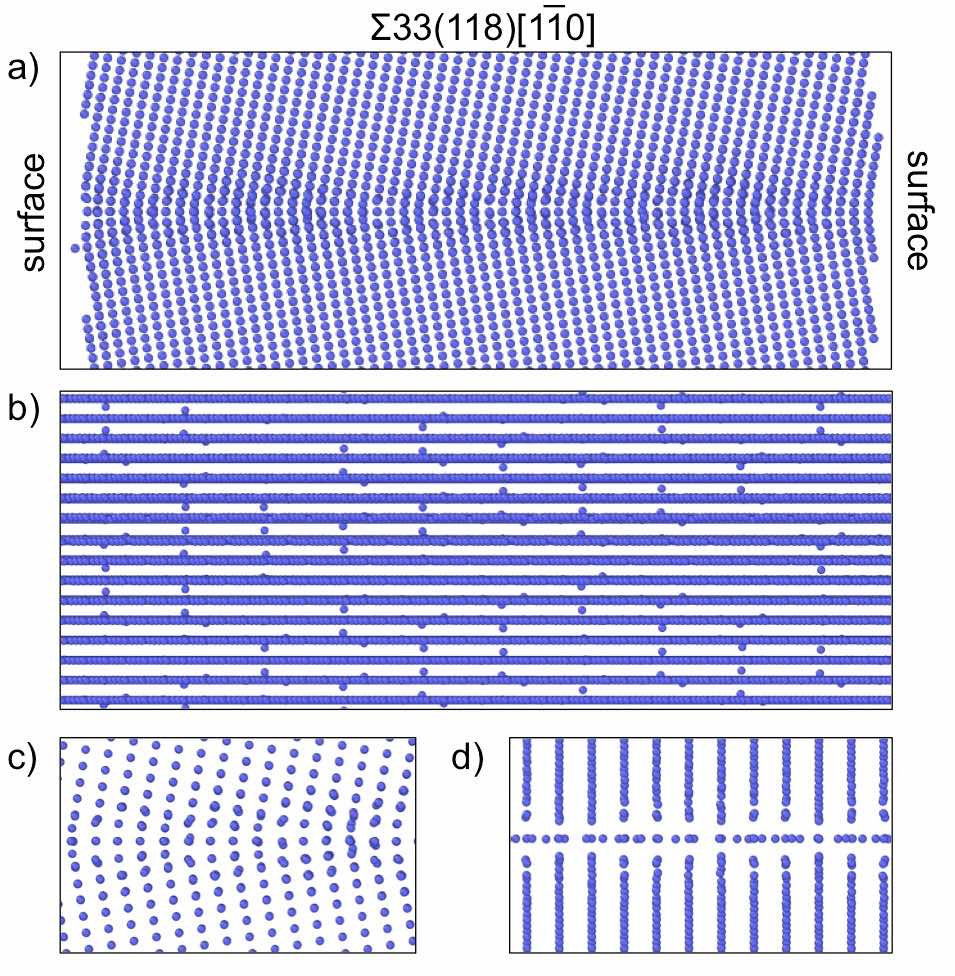} 
\par\end{centering}

\protect\protect\protect\caption{Equilibrium high-temperature structure of the $\Sigma33(118)[1\overline{1}0]$
($\theta=20.1^{\circ}$) GB. (a) The bicrystal terminated at an open
surface was annealed at 2500~K for several tens of nanoseconds (View
1). The open surface enables variation of atomic density. (b) View
3 of the simulation block showing the arrangement of atoms within
the boundary plane (top view). (c) and (d) The zoomed-in views 1 and
2 of the equilibrated boundary structure. The high-temperature GB
structure is different from the $\gamma$-surface constructed boundary,
but matches the prediction of EGCS calculations. \label{fig:HighT-Marin-20}}
\end{figure}

\begin{figure}
\includegraphics[width=1\textwidth]{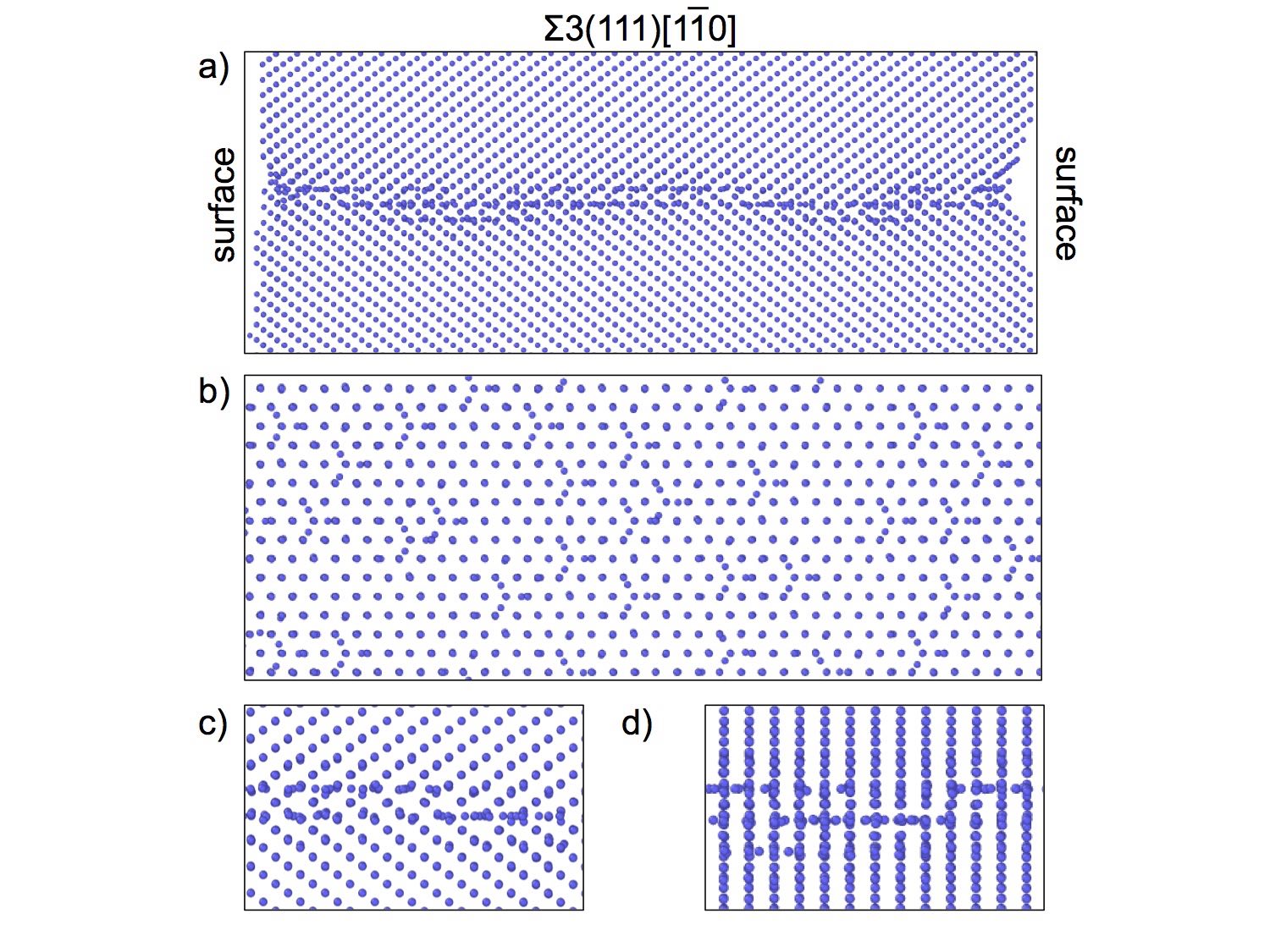}

\protect\protect\protect\caption{Equilibrium high-temperature structure of the $\Sigma3(111)[1\overline{1}0]$
($\theta=109.5^{\circ}$) GB. (a) The bicrystal terminated at an open
surface was annealed at 2500~K for several tens of nanoseconds (View
1). The open surface enables variation of atomic density. (b) View
3 of the simulation block showing the arrangement of atoms within
the boundary plane (top view). (c) and (d) The zoomed-in views 1 and
2 of the equilibrated boundary structure. The high-temperature GB
structure is different from the $\gamma$-surface constructed boundary,
but matches the prediction of EGCS calculations. \label{fig:HighT-Marin-20-1}}
\end{figure}

\begin{figure}
\includegraphics[width=1\textwidth]{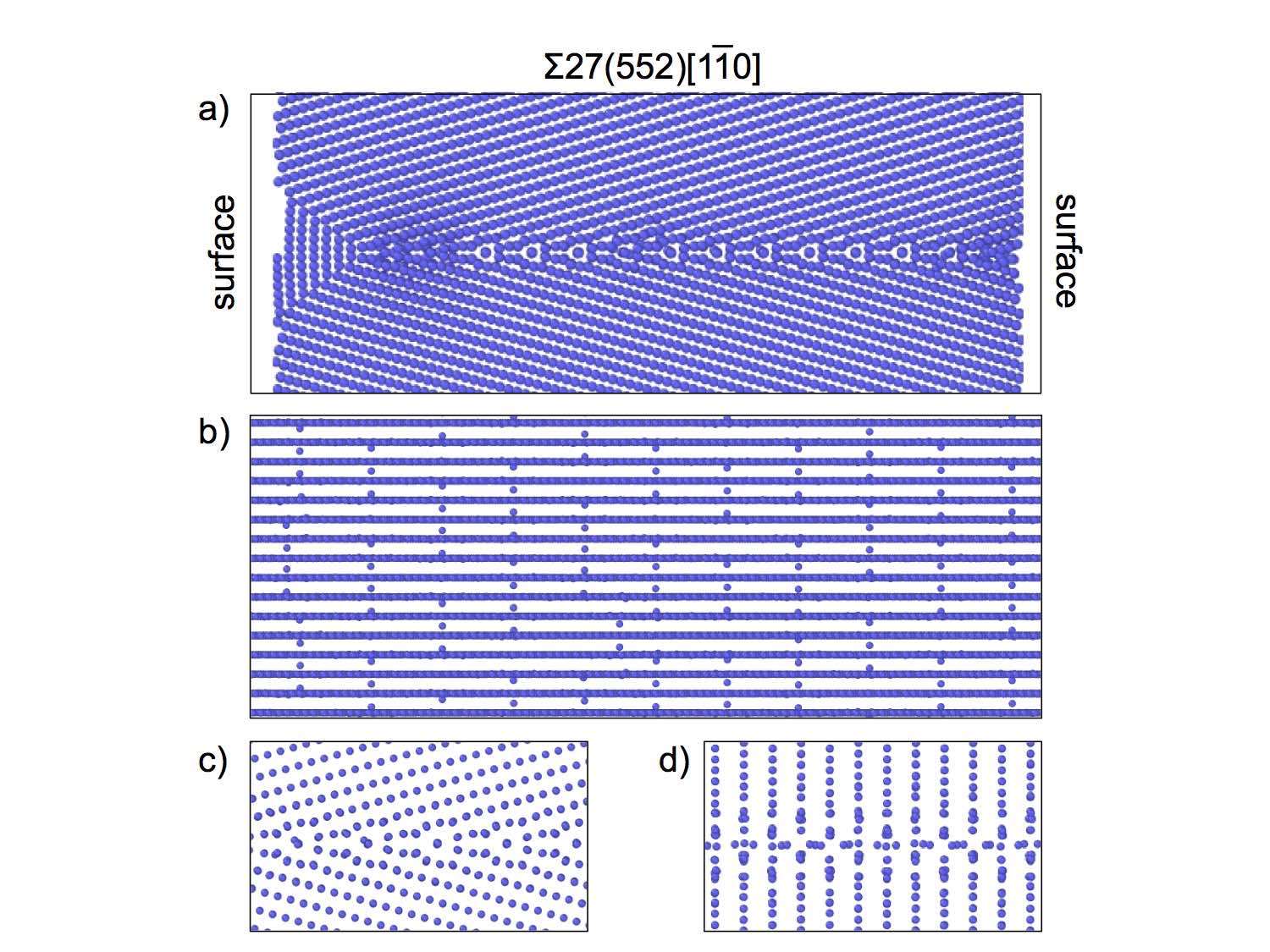}

\protect\protect\protect\caption{Equilibrium high-temperature structure of the $\Sigma27(552)[1\overline{1}0]$
($\theta=148.4^{\circ}$) GB. (a) The bicrystal terminated at an open
surface was annealed at 2500~K for several tens of nanoseconds (View
1). The open surface enables variation of atomic density. b) View
3 of the simulation block showing the arrangement of atoms within
the boundary plane (top view). (c) and (d) The zoomed-in views 1 and
2 of the equilibrated boundary structure. The high-temperature GB
structure is different from the $\gamma$-surface constructed boundary,
but matches the prediction of EGCS calculations. \label{fig:HighT-Marin-148deg}}
\end{figure}

Very different behavior, but consistent with the results of the evolutionary
search at 0~K, was found for the $[110]$ tilt boundaries. Figs.~\ref{fig:HighT-Marin-20},
\ref{fig:HighT-Marin-20-1} and \ref{fig:HighT-Marin-148deg} illustrate
the equilibrium structures of the $\Sigma33(118)[1\overline{1}0]$
$(\theta=20.1^{\circ})$, $\Sigma3(111)[1\overline{1}0]$ $(\theta=109.5^{\circ})$
and $\Sigma27(552)[1\overline{1}0]$ $(\theta=148.1^{\circ})$ tilt
GBs after 200 ns anneals at 2500~K. In all three cases the initial
configurations were generated by the $\gamma$-surface approach. For
all three boundaries these initial structures transformed to the new
configurations during the simulation, confirming the EGCS predictions.
The transformations were accompanied by changes in the atomic density
of GBs. The extra atoms necessary to form the new structures were
supplied by GB diffusion from the open surfaces. The b panels of Fig.~\ref{fig:HighT-Marin-20}-\ref{fig:HighT-Marin-148deg}
correspond to view 3 and show the occupation of the interstitial sites
within the GB plane. This feature of the high-temperature GB phases
is common to all three boundaries and is not characteristic of the
conventional structures generated by the $\gamma$-surface approach.

Fig.~\ref{fig:HighT-Marin-20} reveals that the high-temperature
structure of $\Sigma33(118)[1\overline{1}0]$ $(\theta=20.1^{\circ})$
GB dislocations is more compact than that of the $\gamma$-surface
GB structure. Fig.~\ref{fig:HighT-Marin-20}(c and d) illustrates
closer views of the structure with the tilt axis normal and parallel
to the plane of the figure, respectively. The interstitial columns
in Fig.~\ref{fig:HighT-Marin-20}b (view 3) coincide with the positions
of individual dislocations. Notice that the pattern of the occupied
interstitial sites varies in different dislocations, suggesting that
multiple equivalent sites exist. The interstitial pattern in some
regions of the boundary perfectly matches the structure generated
by the evolutionary algorithm at 0~K, shown in the right-hand panel
of Fig.~\ref{fig:GB-phases-20}(c and d).

Fig.~\ref{fig:HighT-Marin-20-1} illustrates the high-temperature
structure of the $\Sigma3(111)[1\overline{1}0]$ $(\theta=109.5^{\circ})$
GB. The interstitial pattern (view 3) in panel b is very similar to
the pattern generated by the evolutionary search at 0~K. The other
views revealed the complexity of the structure. Fig.~\ref{fig:HighT-Marin-20-1}a
illustrates what appears to be a large number of GB steps. We also
performed an additional simulation with the initial structure $[n]=2/3$
taken from the evolutionary search. The high-temperature simulation
produced a structure with a different GB step pattern; however, the
interstitial pattern was very similar. The $\Sigma3(111)[1\overline{1}0]$
was a relatively challenging boundary to study. It is possible that
longer simulation times and higher temperatures were necessary to
obtain a converged structure of this boundary. On the other hand,
this particular boundary was relatively mobile and traveled by random
walk over distances of several nanometers during the simulation, which
should be sufficient to sample different configurations and adjust
its structure. It is possible that the structures we obtained by high-temperature
simulations do not result from slow kinetics but rather from a property
of this boundary, which is composed of a mixture of several competing
sub-structures.

Fig.~\ref{fig:HighT-Marin-148deg} illustrates the bicrystal with
the $\Sigma27(552)[1\overline{1}0]$ GB modeled with the EAM1 potential
which was annealed at 2500~K for 100 ns. Fig.~\ref{fig:HighT-Marin-148deg}(c
and d) provides closer views of the structure with the tilt axis normal
and parallel to the plane of the figure, respectively. The high-temperature
GB structure matches the $[n]=1/2$ GB phase obtained using the EGCS,
which is illustrated in Fig.~\ref{fig:GB-phases-148M_mult}. The
interstitial pattern shown in Fig.~\ref{fig:HighT-Marin-148deg}b
is similar, but does not match exactly the 0~K patterns shown in
the right-hand panels of Fig.~\ref{fig:GB-phases-148M_mult}. This
again suggests multiple energy-equivalent sites identified at 0~K
by the evolutionary search. The occupation of these sites at finite
temperature is dictated by entropy. The high-temperature structure
also has extra atoms equivalent to half of a (552) plane relative
to the initial configuration obtained using the $\gamma$-surface
approach. The extra atoms diffused inside the GB from the open surface
during the simulation. In addition to the changes in the GB structure,
the surface triple junction on the left-hand side of the figure shows
a chevron reconstruction. Similar reconstructions were previously
observed experimentally by electron microscopy in Au~\citep{PhysRevB.69.172102,PhysRevLett.89.085502}.
The atoms inside the triangular region have perfect bcc structure.
The two boundaries that form the chevron are the $\Sigma3(112)[1\overline{1}0]$
($70.5^{\circ}$) boundaries. Notice that the other surface triple
junction does not undergo a similar reconstruction. Two GB units between
the chevron and the rest of the {[}n{]}=1/2 GB phase have different
structures, which closely resemble the $[n]=0$ structure generated
by the EAM2 potential. While the EAM1 does not predict a low-energy
configuration at this atomic fraction, it is possible that this alternative
structure is stabilized by the mechanical stresses near the triple
junction.

\subsubsection{GB phase coexistence and point defect absorption in simulations with
periodic boundary conditions}

\subsubsection*{{[}100{]}-boundaries}

\begin{figure}
\includegraphics[width=1\textwidth]{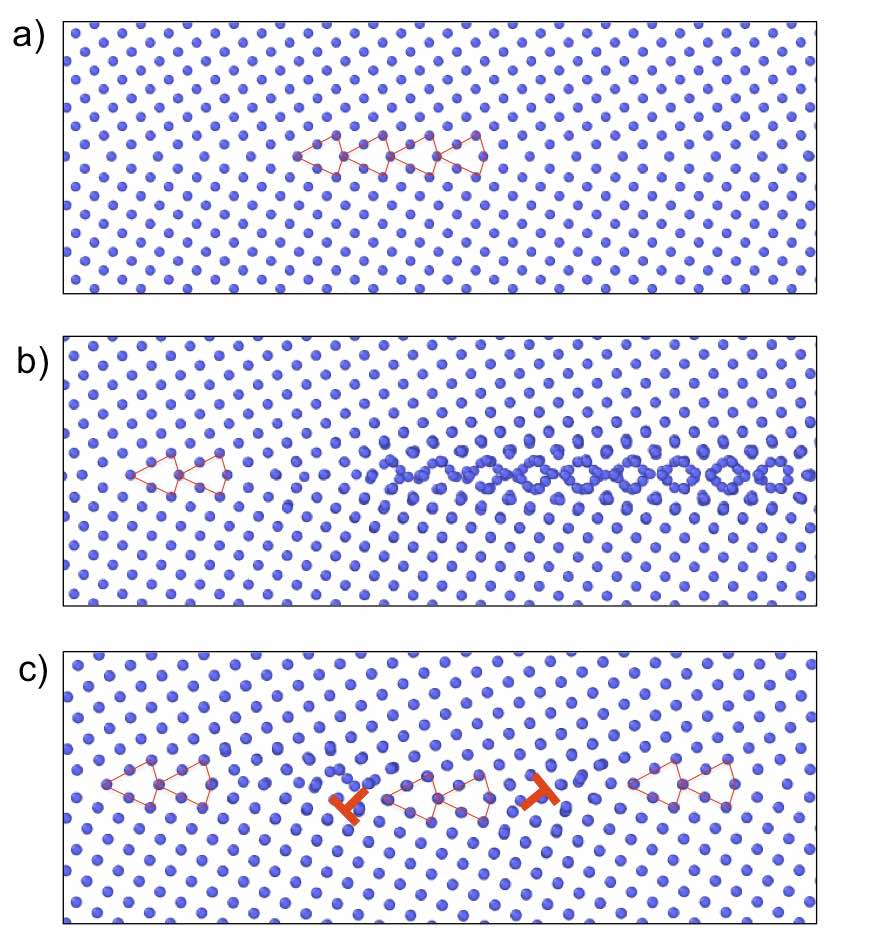}

\protect\protect\protect\caption{(a) Interstitials are introduced into a bicrystal with a perfect $\Sigma5(310)[001]$
GB. The kite-shaped structural units in the GB are outlined in red.
(b) The atoms diffuse to the boundary and get absorbed by locally
forming a metastable ordered GB structure with high energy. (c) At
later times the metastable GB segment transforms into an interstitial
loop at the GB. \label{fig:High310}}
\end{figure}

To observe possible metastable states of the $\Sigma5(310)[001]$
boundary with higher atomic densities, we introduced interstitials
into the bulk lattice just above the GB plane and annealed the blocks
at 2000~K and 2500~K in separate simulations. The periodic boundary
conditions were applied parallel to the boundary plane to eliminate
sinks for the interstitial atoms. At both temperatures, we first observed
formation of an ordered GB structure due to absorption of the interstitials.
Fig.~\ref{fig:High310}b illustrates the two different states of
the boundary, which are similar to the structures observed by Novoselov
and Yunilkin in bcc Mo \citep{Novoselov2016276}. This metastable
configuration exists for almost 100 ns at 2000~K and several tens
of nanoseconds at 2500~K before transforming into an interstitial
loop at the boundary. The final state of the boundary is illustrated
in Fig.~\ref{fig:High310}c. The GB segment confined between the
two GB dislocations is composed of perfect kite-shaped structural
units. These new units appeared out of the metastable GB configuration
demonstrating that the Kite structure of this boundary is very stable
even at this high temperature. The relatively long lifetime of the
metastable high-energy state is probably due to a large barrier of
transformation that involves nucleation of the GB dislocations.

\begin{figure}
\includegraphics[width=1\textwidth]{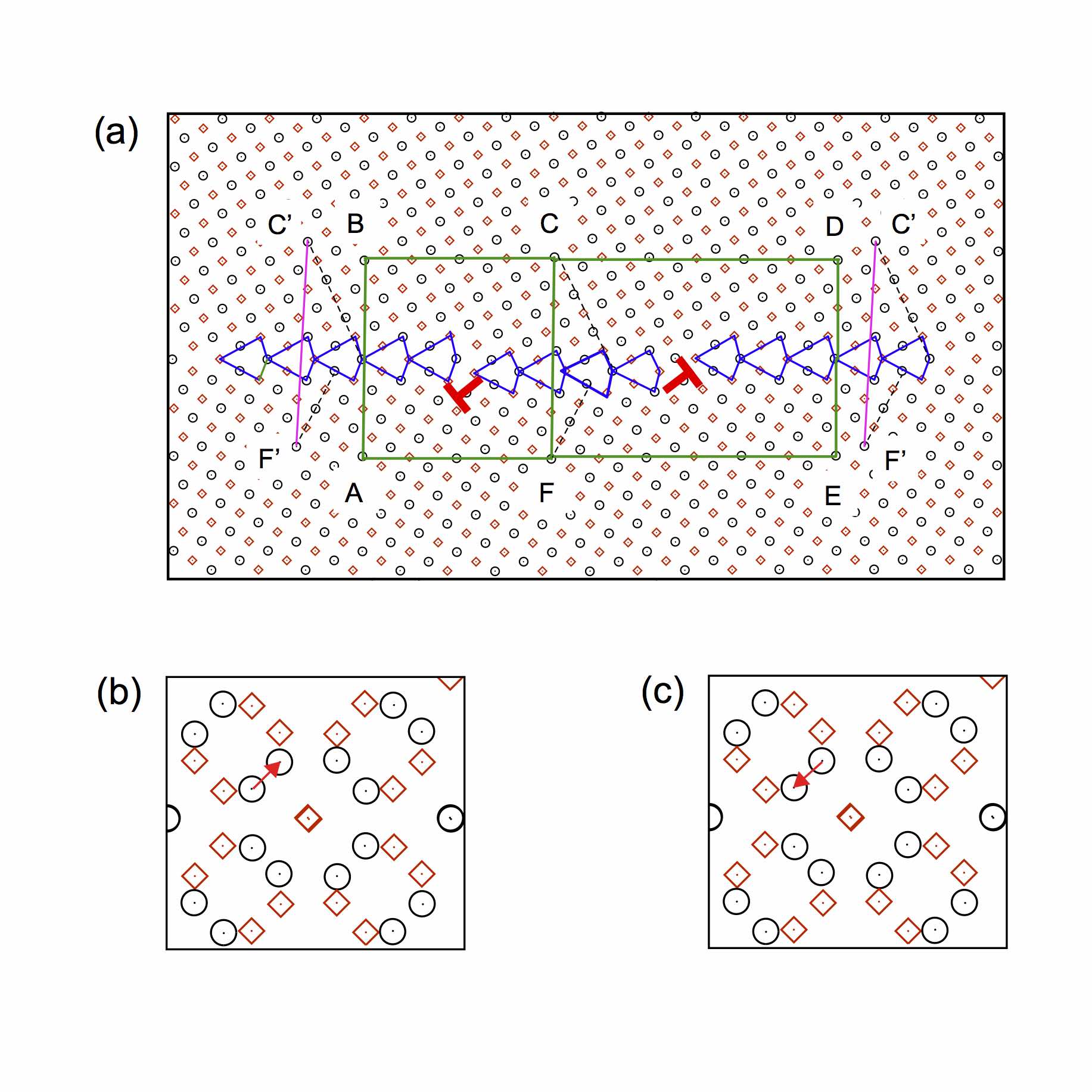}

\protect\protect\protect\caption{(a) An interstitial loop at the $\Sigma5(310)[001]$ GB represented
by two disconnections $(1/10[310]a,1/10[310]a,0)$ and $(-1/10[310]a,1/-10[310]a,0)$
identified by circuits ABCF and FCDE. The ABDE circuit encloses the
entire dislocation loop and as expected has a net zero Burgers vector.
The Burgers vectors of each disconnection is a DCS vector equal to
the sum b) AB and C'F' and c) F'C' and DE vectors, respectively. F'C'
corresponds to FC vector on the reference lattice. \label{fig:Disconnections310-2}}
\end{figure}

To characterize the GB disconnections we constructed closed circuits
ABCF and FCDE around each of the line defects as illustrated in Fig.~\ref{fig:Disconnections310-2}.
The red and black lattice sites are colored according to their position
normal to the plane of the figure. The ABDE circuit connects the four
black lattice sites and encloses the entire GB dislocation loop. The
AB and DE segments cut through identical perfect GB structures and
have the same length. BD and EA segments have the same length as well.
As a result, the total disconnection content of the ABDE circuit is
zero, same as that of a defect-free GB.

To calculate the disconnection content of the ABCF and FCDE circuits,
we find the FC vector that cuts through the middle section of the
GB loop on the reference lattice. The corresponding F'C' vectors are
illustrated in Fig.~\ref{fig:Disconnections310-2}a. Then, we sum
the four vectors A'B', B'C', C'F' and F'A' using their length measured
on the reference stress-free lattice. It is clear that vectors B'C'
and F'A' do not contribute to the disconnection, since they are just
lattice vectors with the same magnitude and opposite signs and cancel
each other. The sum of the other two vectors A'B'+C'F'=-(B'C'+F'A'),
since A'B'C'F' is a closed loop on a disconnection-free bicrystal.
The sum -(B'C'+F'A') of the two lattice vectors that belong to two
different crystals is a DSC vector with components (1/10{[}310{]}a,
1/10{[}310{]}a, 0). Analogous construction identifies -(1/10{[}310{]}a,
1/10{[}310{]}a, 0) Burgers vector for the other disconnection. The
DSC vectors of the two disconnections are illustrated in Fig.~\ref{fig:Disconnections310-2}b
and c, respectively. The non-zero components of the Burgers vectors
normal to the GB plane indicate that the extra materials was accommodated
by an interstitial loop at the boundary. The Burgers circuit analysis
used in this work is somewhat different from the analysis described
in Refs. \citep{Hirth96,Pond03a,HIRTH2013749}. However, it can be
shown to be equivalent for the case when the circuit cuts two identical
GB segments, which is the case here.

\subsubsection*{{[}110{]}-boundaries}

\begin{figure}
\includegraphics[width=1\textwidth]{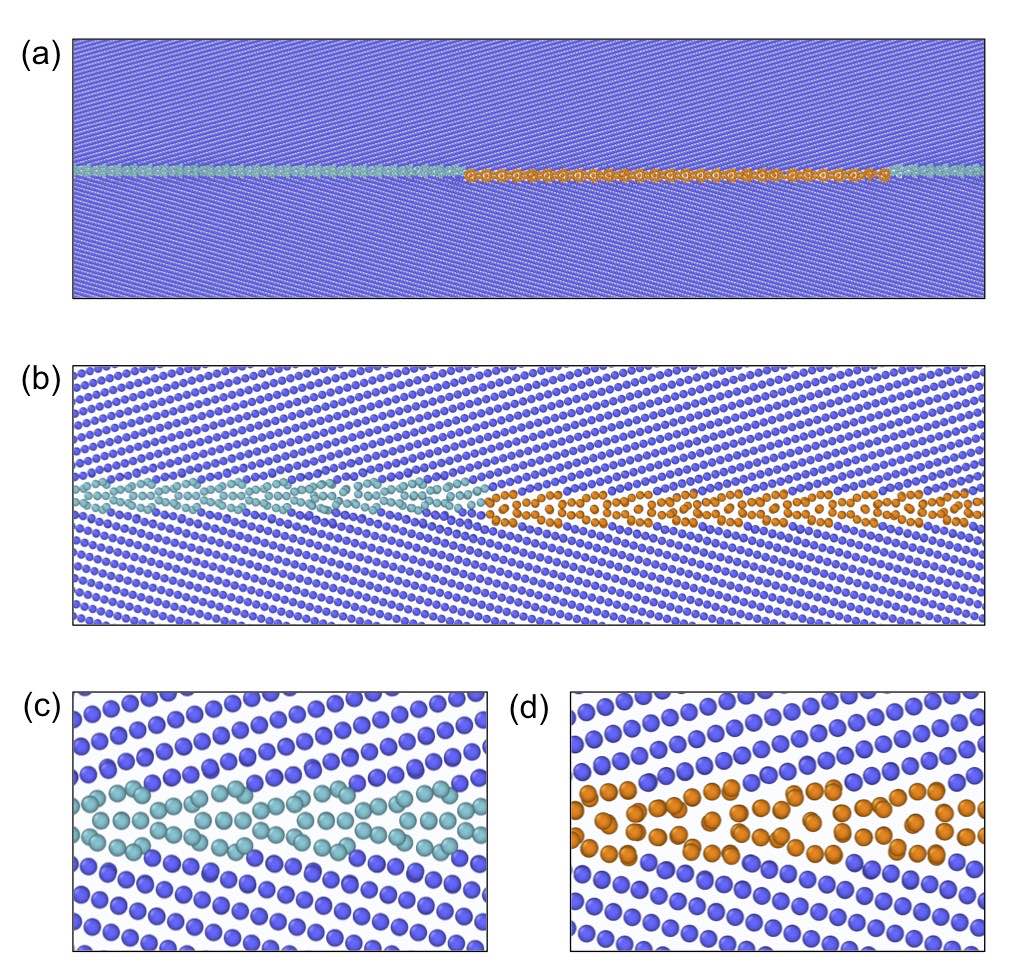}

\protect\protect\protect\caption{(a) Equilibrium coexistence of the {[}n{]}=0 and {[}n{]}=1/2 GB phases
of the $\Sigma27(552)[0\overline{1}1]$ GB in a closed system with
periodic boundary conditions at 1500~K. The two GB phase coexistence
is implemented by introducing interstitials into perfect {[}n{]}=0
GB phase predicted by the evolutionary search. The interstitials are
absorbed when about a half of the boundary transforms into the {[}n{]}=1/2
GB phase, also predicted by the evolutionary search. The size of each
GB phase is about 25 nm in the $x$ direction. (b) Two GB phases meet
along a line defect that spans the periodic length of the simulation
block. (c) and (d ) are zoomed-in views of the two GB phases. The
GB atoms are identified according to common neighbor analysis~\citep{0965-0393-18-1-015012}.
\label{fig:148_coexist}}
\end{figure}

To test the response of the boundary with multiple GB phases to the
changes in the atomic density {[}n{]}, we performed MD simulations
of the $\Sigma27(552)[1\overline{1}0]$ GB with the EAM2 potential.
When periodic boundary conditions are applied both GB phases $[n]=0$
and $[n]=1/2$ are stable at high temperature. The constraint of the
constant number of atoms insures that one structure does not transform
into another during the simulation. We used the $[n]=0$ structure
as the initial configuration and inserted extra atoms in the bulk
lattice just above the GB plane. The interstitials triggered a nucleation
of the $[n]=1/2$ GB phase. The areal fraction of the new GB phase
was dictated by the number of extra atoms introduced. Fig.~\ref{fig:148_coexist}a
illustrated the structure of the boundary with two GB phases at 1500
K. The two phases are colored in orange and green in Fig.~\ref{fig:148_coexist}b.
They are separated by a GB phase junction, a line defect that spans
the periodic dimension normal to the plane of the figure. Fig.~\ref{fig:148_coexist}(c
and d) shows zoomed in views of the two GB structures. In contrast
to the $\Sigma5(310)[001]$ boundary, no other transformations occurred
in this simulation: the heterogeneous boundary with the two different
GB phases coexisting was the final state of the simulation.

The stable equilibrium is established because the boundary is isolated
from the sources and sinks of atoms. During the simulation GB atoms
diffuse to establish an equilibrium concentration of vacancies or
interstitials in the two different GB structures. The positions of
the GB phase junctions dynamically fluctuate during the coexistence
simulation, so that a small portion of one boundary constantly attempts
to transform into the other. During such a transformation extra atoms
are produced or absorbed, because the GBs have different densities
{[}n{]}. After each fluctuation, these extra atoms or vacancies are
redistributed among the two GB structures by diffusion and change
their free energy in a way to prevent further transformation. Thus,
the equilibrium in such a closed system is stable. In principle, the
extra GB atoms could escape to the surfaces through the bulk. However,
the equilibrium concentration of interstitials at these temperatures
is so low that such a transformation is very unlikely to observe on
the MD time scale.

This type of equilibrium is unique to solid systems because the solid
lattice and varying number of atoms provide the system with an additional
thermodynamic degree of freedom~\citep{Larche_Cahn_78,doi:10.1063/1.448644,Voorhees20041}.
Indeed, according to Gibbs phase rule in an elemental system at a
fixed pressure two GB phases should be able to coexist only at one
temperature~\citep{Frolov:2015ab}. MD simulations of GB phase transitions
in elemental systems follow this prediction when the boundary is connected
to source/sink of atoms~\citep{Frolov2013}. On the other hand, the
GB phase coexistence in a closed system such as illustrated in Fig.~\ref{fig:148_coexist}
persists in a range of temperatures. In this work we simulated two-phase
coexistence at 1500 K, 1800 K and 2000 K. The temperature changes
the number of atoms {[}n{]} in each of the phases: the equilibrium
concentration of vacancies and interstitials present in the coexisting
GB phases. At 2500 K the {[}n{]}=1/2 GB phase to transformed into
{[}n{]}=0. Here we label the two GB phases by referring to their atomic
density at 0K, which changes with temperature. During the transformation
the extra atoms are accommodated as defects of the {[}n{]}=0 GB phase,
which apparently become energetically inexpensive at this high temperature.
The solubility of defects in each GB phase became such that crossing
the coexistence line became possible even in a closed system.

\subsubsection*{Nucleation and transformation of small GB islands}

\begin{figure}
\includegraphics[width=1\textwidth]{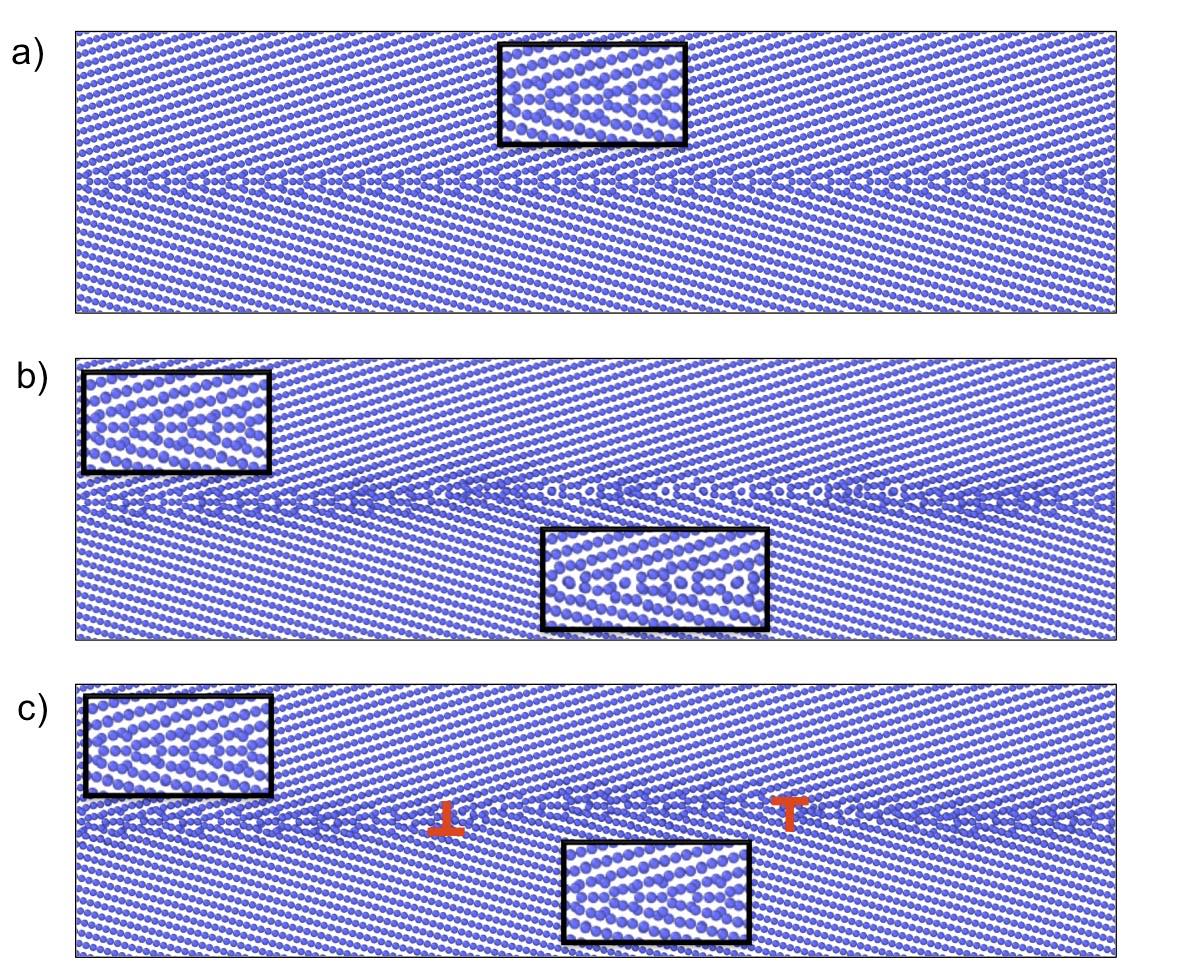}

\protect\protect\protect\caption{(a) A smaller number of interstitials is introduced into a bicrystal
with the $\Sigma27(552)[0\overline{1}1]$ GB in an isothermal simulation
at 2000~K. The initial structure corresponds to the $[n]=0$ phase
predicted by the evolutionary search with the EAM2 potential. After
interstitial atoms are introduced in the bulk part of the upper crystal
just above the GB, they quickly diffuse to the boundary core. There
the interstitials are absorbed when a relatively small portion of
the boundary transforms into {[}n{]}=1/2 GB phase. The size of the
{[}n{]}=1/2 phase is about 6 nm in the $x$ direction. (b) During
the subsequent 50-ns-long isothermal simulation both GB phases coexist
in equilibrium while exchanging atoms which diffuse along the boundary.
The two different GB phases are shown in different colors. The coloring
of the {[}n{]}=1/2 structure is from a common neighbor analysis. (c)
After 50 ns {[}n{]}=1/2 phase transforms into an interstitial loop.
The simulation suggests that the stability of the heterogeneous GB
structure with respect to nucleation of an interstitial loop may be
size dependent. \citep{0965-0393-18-1-015012}. \label{fig:High_146_interst_small}}
\end{figure}

We find that even in the case when the different GB phases have close
energies the stability of the heterogeneous GB structure may be size
dependent. When a smaller number of interstitials is introduced and
only a few structural units of the {[}n{]}=1/2 boundary are formed,
the small islands of the new GB phase eventually transform into the
{[}n{]}=0 structure at 2000 K. Fig.~\ref{fig:High_146_interst_small}(a
and b) illustrate the initial homogeneous {[}n{]}=0 GB structure and
the heterogeneous GB structure after the interstitials were absorbed,
respectively. The mechanism is somewhat analogous to the $\Sigma5(310)[001]$
boundary, in this two-step process a small island of the {[}n{]}=1/2
phase nucleates first (Fig.~\ref{fig:High_146_interst_small}b) and
after several tens of nanoseconds it transforms into a different structure
closely resembling the {[}n{]}=0 GB phase. The defected structure
is separated from the original boundary by two GB disconnections.

\begin{figure}
\includegraphics[width=1\textwidth]{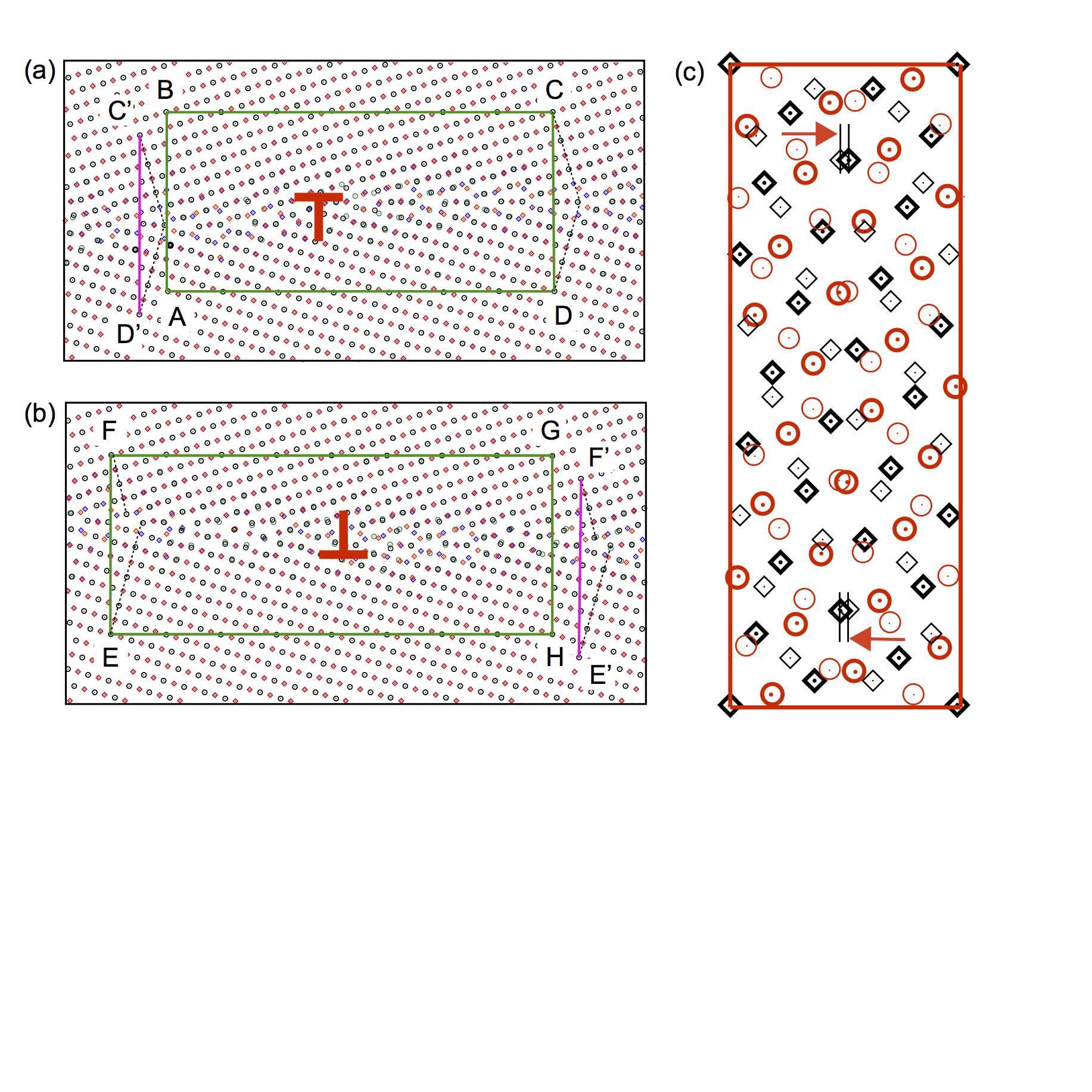}

\protect\protect\caption{Disconnections at the $\Sigma27(552)[0\overline{1}1]$ GB after an
island of the {[}n{]}=1/2 GB phase transforms into the n={[}0{]} GB
phase at 2000~K. The circuits (b) ABCD and (c) EFGH identify two
disconnections $[1/27[115]a/2,0,0]$ and $[-1/27[115]a/2,0,0]$. c)
Burgers vectors of the two disconnections shown as vectors of the
DSC lattice. \label{fig:Disconnections-148}}
\end{figure}

We analyzed the two disconnections by constructing two closed circuits
around the line defects as illustrated in Fig.~\ref{fig:Disconnections-148}(a
and b). The CD and EF vectors cut through the transformed section
of the boundary. We note that while the image of the boundary projected
on the plane of the screen matches the original {[}n{]}=0 structure,
the examination of the atomic positions within the plane revealed
that the transformed GB segment is significantly different and appears
to have defects and even small sections of the {[}n{]}=1/2 phase.
Two similar variants with different densities {[}n{]}=0 and {[}n{]}=1/3
were recently demonstrated in the same $\Sigma27$ boundary in Ta
\citep{Frolov2018Nanoscale}. It is possible, that a similar situation
occurs in our simulations in tungsten. Neglecting the difference between
two possible variants, the vectors of the two circuits were summed
following the procedure described earlier for the $\Sigma5(310)[001]$
GB and identified disconnections with {[}1/27{[}115{]}a/2,0,0{]} and
{[}-1/27{[}115{]}a/2,0,0{]} Burgers vectors. The zero component normal
to the plane of the GB suggests that the extra atoms were accommodated
as defects of the boundary and not as a GB dislocation loop.

\section{Discussion and conclusions\label{sec:Discussion}}

In this work we studied {[}001{]} and $[110]$ symmetric tilt GBs
in bcc tungsten. These boundaries have been studied previously by
atomistic simulations with empirical potentials and DFT calculations
in several bcc materials including W, Mo and Fe. In these studies
the GBs were generated using the common $\gamma$-surface method that
performs limited sampling of GB structure and does not attempt to
add or remove atoms from the GB core.

In the current work, we generate the boundary structures using the
new evolutionary approach \citep{doi:10.1063/1.2210932,Zhu2018}.
This algorithm samples a diverse range of different structures, optimizes
GB atomic density and searches for larger area reconstructions. The
re-examination of the structure of these symmetric tilt GBs was motivated
by recent work in fcc boundaries, that demonstrated that in several
model systems kite-shaped structural units have limited stability
and alternative GB structures were predicted to be the ground states
at 0~K and finite temperature \citep{Frolov2013,Zhu2018}.

\subsection{{[}001{]} tilt boundaries}

For the {[}001{]} symmetric tilt GBs studied in this work, the $\gamma$-surface
approach predicts configurations composed of kite-shaped structural
units in agreement with previous studies. Our grand canonical search
confirmed these structures to be the ground state for two representative
high-angle high-energy GBs. Thus, contrary to {[}001{]} tilt boundaries
in fcc models of Cu, Ag, Au and Ni, in bcc W the kite-shaped structural
units are stable. We find alternative ordered metastable structures
with higher atomic density by loading the ground state with interstitials.
In these structures the extra atoms occupy interstitial positions
within the GB plane located between the (001) planes of the abutting
crystals. Similar structures were reported earlier in bcc Mo \citep{Novoselov2016276}.
Our high-temperature MD simulations indicate that these denser states
are stable against dissolution in the parent Kite structure even in
the presence of rapid GB diffusion and survive at high temperature
for relatively long time on the MD time scale. Their lifetime depends
on the temperature. However, the energy of these states is still significantly
higher, and at high temperature we observe a transformation into the
Kite phase, which results in the formation of an interstitial loop
at the GB. This transformation confirms the stability of the Kite
structure even at high temperature. These modeling results are consistent
with experimental observations of GB structure in other bcc metals.
For example, in Mo the kite-shaped GB structure of the $\Sigma5(310)[001]$
was directly observed by high-resolution transmission electron microscopy
\citep{CAMPBELL19993977,doi:10.1080/01418610208240038}. In Fe, a
study of an asymmetric {[}001{]} boundary demonstrated faceting into
$\Sigma5(310)[001]$ and $\Sigma5(210)[001]$ symmetric tilt boundaries
with perfect Kite structures \citep{Medlin2017383,doi:10.1063/1.4954066}.
The atomic structure of the faceted boundary was observed by high-resolution
electron microscopy and simulated with molecular dynamics. We conclude
that in these {[}001{]} symmetric tilt GBs studied in elemental tungsten,
the $\gamma$-surface method is likely to be sufficient to generate
the GB structure at 0~K and finite temperature. The situation may
be different in doped systems. A recent study of the $\Sigma5(210)[001]$
Mo GB demonstrated a first-order structural transition induced by
segregation of Ni \citep{PhysRevLett.120.085702}. Similar transitions
have been demonstrated by atomistic simulations in other systems \citep{PhysRevB.92.020103,OBrien2018}.

\subsection{{[}110{]} tilt boundaries}

For the majority of the {[}110{]} symmetric tilt boundaries studied
in this work, which includes both high-angle and low-angle GBs, the
EGCS method revealed new ground states and multiple GB phases, demonstrating
that the $\gamma$-surface method is insufficient to predict the correct
GB structure in these model systems. The novel GB structures cannot
be described by the conventional GB structural units and they share
several common features. Most of them are composed of a number of
atoms incompatible with the number of atoms in the lattice planes
of the abutting crystals. To obtain these structures extra atoms must
be inserted into the GB core. In these structures the atoms occupy
interstitial positions within the boundary plane located in between
the misoriented (110) planes. The evolutionary search generated many
configurations degenerate in energy, characterized by different occupation
of these interstitial positions within the boundary. The multiplicity
of these states may contribute to configurational entropy and affect
the stability of these structures at high temperature. It is well
known that the $\gamma$-surface approach can also generate distinct
GB structures with the same energy, corresponding to different grain
translation vectors. For the new structures generated by the EGCS,
the multiple energetically degenerate states are related by permutation
of interstitial atoms without changing the grain translation vector.
Finally, most of the the new structures have an irreducible unit larger
than the periodic units of the CSL lattice. The GB reconstructions
with different dimensions often had very similar energies.

The only two {[}110{]} boundaries that did not share these properties
were the $\Sigma3(112)[1\overline{1}0]$ and the $\Sigma11(332)[1\overline{1}0]$
at $\theta=66.22^{\circ}$ and $\theta=129.5^{\circ}$, respectively.
These correspond to two energy cusps as a function of the misorientation
angle. The $\Sigma3(112)[1\overline{1}0]$ has the lowest energy because
of its almost bulk-like structure, so it was not surprising that the
evolutionary search did not find alternative low-energy configurations.
The second cusp at $\theta=129.5^{\circ}$ has a noticeably higher
energy, but was also identified as a stable ground state by the evolutionary
search in agreement with the $\gamma$-surface method.

In some cases the new ground states generated by the EGCS algorithm
had energies significantly lower than those generated by the conventional
methodology, while in other cases the energies were nearly identical.
For example, in the case of the $\Sigma27(552)[1\overline{1}0]$ boundary,
the energy was reduced by 7 - 12\% depending on the potential. In
all boundaries with multiple distinct phases the energy difference
was very small, within a few percent. While the energy reduction obtained
by the advanced search was modest in some cases, the properties of
different GB phases may differ significantly. For example in fcc Cu,
the simulations demonstrated a strong effect of the transitions on
self and impurity diffusion~\citep{PhysRevB.92.020103,Frolov2013,Frolov2013PRL},
segregation~\citep{PhysRevB.92.020103} as well as GB migration and
shear strength~\citep{doi:10.1063/1.4880715}. A recent study investigated
coupled motion of two {[}110{]} symmetric tilt boundaries in bcc iron
and demonstarted abrupt changes in GB migration and shear stress with
increasing temperature \citep{YIN2018141}. These results are consistent
with multiple GB phases and GB phase transitions demontrated in our
study for the same family of symmetric tilt boundaries in a different
bcc metal. 

We find that overall the predictions of the two potentials EAM1 and
EAM2 are consistent. Both potentials predict similar trends for the
GB energy as a function of the misorientation angle. For some boundaries
such as the $\Sigma33(118)[1\overline{1}0]$ $(\theta=20.1^{\circ})$
the two potentials predicted different structures within the $\gamma$-surface
approach. Prior DFT calculations reported that the same GB can have
these different structures in different bcc materials \citep{swissW2016}.
However, the evolutionary search for the $\Sigma33(118)[1\overline{1}0]$
boundary predicted the same ground state at {[}n{]}=1/3 with both
potentials. This example suggests that in some cases the discrepancy
in the structure predicted by different models may be an artifact
of the $\gamma$-surface approach and not the issue of the force field.

\subsection{GB structures and transitions at finite temperature}

The multiplicity of distinct GB structures with very close energies
found at 0~K motivated further investigation of the finite-temperature
GB structure. In this work we performed MD simulations at high temperature
with the GBs terminated at open surfaces. The surfaces act as sources
and sinks of atoms. These simulations demonstrated transformations
from the {[}n{]}=0 $\gamma$-surface generated structures to the structures
predicted by the evolutionary with other atomic densities. Thus, despite
the close energetics at 0~K, we found that the non-conventional GB
structures become more stable at finite temperature. In fact, with
the exception of two boundaries at $\theta=66.22^{\circ}$ and $\theta=129.5^{\circ}$,
the structures generated by the $\gamma$-surface approach do not
represent the finite-temperature structure of the {[}110{]} symmetric
tilt boundaries studied.

The simulated transitions suggest that finding the lowest energy configurations
and 0~K may not be sufficient to predict the structure and properties
of GBs at finite temperature. In the current study the high-temperature
structures were generated by the evolutionary search at 0 K and coincided
with the energy minima as a function of the atomic density {[}n{]}.
In general, this should not be expected. In our investigation of the
{[}100{]} symmetric tilt fcc Cu boundaries we demonstrated that the
high-temperature state does not correspond to GB energy minima~\citep{Zhu2018},
and a more sophisticated analysis is required to extract potential
high-temperature structures from the results of the 0~K structure
search. Specifically, we proposed a clustering procedure that groups
individual structures generated by the evolutionary algorithm into
GB phases. Although, we performed the grand canonical structure search
for only a small subset of the boundaries, it likely that many other
{[}110{]} boundaries exhibit unusual structures and multiple phases.
A detailed investigation metastable structures and possible structural
trends of the {[}110{]} tilt boundaries is left to future work.

High-temperature simulations with periodic boundary conditions and
added point defects demonstrated nucleation of a second GB phase with
different atomic density. Defect induced GB transitions have been
demonstrated previously in Cu \citep{Frolov2013} and W \citep{Frolov2018Nanoscale}.
Atomistic simulations also demonstrated that cracks and voids can
be healed through a formation of a new boundary segment with a different
atomic density \citep{ARAMFARD2018304}. For this simulation we selected
the $\Sigma27(552)[1\overline{1}0]$ modeled with the EAM2 potential,
because the boundary exhibits two different structures with the same
energy at 0~K and very different atomic densities of {[}n{]}=0 and
{[}n{]}=1/2. The simulations revealed that after the nucleation of
the {[}n{]}=1/2 phase, the two structures can coexist while exchanging
atoms through GB diffusion. The coexistence simulations confirm that
the structures represent two phases of this boundary and are not just
mechanically stable configurations at 0~K. In some simulations we
observed that after about 100 ns of coexistence the small secondary
phase transforms into an interstitial loop at the boundary. This behavior
is exactly analogous to the two-step nucleation of the interstitial
loop at the $\Sigma5(310)[001]$ boundary, when the formation of a
high-energy metastable GB structure induced by interstitials is followed
by nucleation of GB dislocations. In the $\Sigma27(552)[1\overline{1}0]$
case, however, both GB phases have the same energy, and we speculate
that the transition is driven by elastic interactions between the
GB phase junctions. These line defects separate different GB phases
and are likely to have dislocation character~\citep{Hirth96}. A
heterogeneous boundary with a secondary phase and a homogeneous boundary
with an interstitial loop represent two competing states of the boundary
after it absorbs point defects. Our simulations suggest that the absorption
by nucleation of a secondary GB phase is kinetically preferred, while
the loop formation is more energetically favorable for some systems
studied. Two-step nucleation is a well known phenomenon in bulk materials
and is often observed during solidification \citep{doi:10.1021/ar800217x}.
Here we extended it to process at grain boundaries, where new interface
specific factors may play an important role. For example, the stability
of the heterogeneous boundaries with respect to loop nucleation or
a formation of other GB phase should be influenced by elastic interactions
in these systems, which are likely to be size dependent. The existing
fluid-like treatments of GB phases neglect elastic effects \citep{Frolov:2015ab,Tang06}.
The simulations motivate the development of a nucleation model that
takes these elastic interactions into account.

\section*{Acknowledgment }

This work was performed under the auspices of the U.S. Department
of Energy (DOE) by Lawrence Livermore National Laboratory under contract
DE-AC52-07NA27344. This material is based upon work supported by the
U.S. DOE, Office of Science, Office of Fusion Energy Sciences. The
work was supported by the Laboratory Directed Research and Development
Program at LLNL, project 17-LW-012. We acknowledge the use of LC computing
resources. Work in UNLV is supported by the National Nuclear Security
Administration under the Stewardship Science Academic Alliances program
through DOE Cooperative Agreement DE-NA0001982.

\clearpage{}

\begin{thebibliography}{10}
\newcommand{\enquote}[1]{`#1'}

\bibitem{Balluffi95}
Sutton, A.~P. and Balluffi, R.~W., {\em Interfaces in Crystalline Materials\/},
  Clarendon Press, Oxford, 1995.

\bibitem{WEI20064079}
Wei, Q., Zhang, H., Schuster, B., Ramesh, K., Valiev, R., Kecskes, L., Dowding,
  R., Magness, L. and Cho, K., {\em Acta Materialia\/}, 2006, {\bf 54}, 4079 .

\bibitem{Chookajorn951}
Chookajorn, T., Murdoch, H.~A. and Schuh, C.~A., {\em Science\/}, 2012, {\bf
  337}, 951.

\bibitem{El-Atwani:2017aa}
El-Atwani, O., Nathaniel, J.~E., Leff, A.~C., Hattar, K. and Taheri, M.~L.,
  {\em Scientific Reports\/}, 2017, {\bf 7}, 1836.

\bibitem{ElAtwani201568}
El-Atwani, O., Suslova, A., Novakowski, T., Hattar, K., Efe, M., Harilal, S.
  and Hassanein, A., {\em Materials Characterization\/}, 2015, {\bf 99}, 68 .

\bibitem{Bai26032010}
Bai, X.-M., Voter, A.~F., Hoagland, R.~G., Nastasi, M. and Uberuaga, B.~P.,
  {\em Science\/}, 2010, {\bf 327}, 1631.

\bibitem{Uberuaga:2015aa}
Uberuaga, B.~P., Vernon, L.~J., Martinez, E. and Voter, A.~F., {\em Scientific
  Reports\/}, 2015, {\bf 5}, 9095.

\bibitem{doi:10.1146/annurev-matsci-070813-113627}
Zinkle, S. and Snead, L., {\em Annual Review of Materials Research\/}, 2014,
  {\bf 44}, 241.

\bibitem{0029-5515-47-11-014}
Maisonnier, D., Campbell, D., Cook, I., Pace, L.~D., Giancarli, L., Hayward,
  J., Puma, A.~L., Medrano, M., Norajitra, P., Roccella, M., Sardain, P., Tran,
  M. and Ward, D., {\em Nuclear Fusion\/}, 2007, {\bf 47}, 1524.

\bibitem{Mutoh95}
Mutoh, Y., Ichikawa, K., Nagata, K. and Takeuchi, M., {\em J. of Mat. Sci.\/},
  1995, {\bf 30}, 770.

\bibitem{Cantwell20141}
Cantwell, P.~R., Tang, M., Dillon, S.~J., Luo, J., Rohrer, G.~S. and Harmer,
  M.~P., {\em Acta Materialia\/}, 2014, {\bf 62}, 1 .

\bibitem{Dillon20076208}
Dillon, S.~J., Tang, M., Carter, W.~C. and Harmer, M.~P., {\em Acta Mater.\/},
  2007, {\bf 55}, 6208.

\bibitem{Luo99}
Luo, J., Wang, H. and Chiang, Y.-M., {\em Journal of the American Ceramic
  Society\/}, 1999, {\bf 82}, 916.

\bibitem{Luo23092011}
Luo, J., Cheng, H., Asl, K.~M., Kiely, C.~J. and Harmer, M.~P., {\em
  Science\/}, 2011, {\bf 333}, 1730.

\bibitem{Baram08042011}
Baram, M., Chatain, D. and Kaplan, W.~D., {\em Science\/}, 2011, {\bf 332},
  206.

\bibitem{Harmer08042011}
Harmer, M.~P., {\em Science\/}, 2011, {\bf 332}, 182.

\bibitem{Rheinheimer201568}
Rheinheimer, W. and Hoffmann, M.~J., {\em Scripta Materialia\/}, 2015, {\bf
  101}, 68 .

\bibitem{Pan201791}
Pan, Z. and Rupert, T.~J., {\em Scripta Materialia\/}, 2017, {\bf 130}, 91 .

\bibitem{SCHULER2017196}
Schuler, J.~D. and Rupert, T.~J., {\em Acta Materialia\/}, 2017, {\bf 140}, 196
  .

\bibitem{Tang06}
Tang, M., Carter, W.~C. and Cannon, R.~M., {\em Phys. Rev. {\rm B}\/}, 2006,
  {\bf 73}, 024102.

\bibitem{Tang06b}
Tang, M., Carter, W.~C. and Cannon, R.~M., {\em Phys. Rev. Lett.\/}, 2006, {\bf
  97}, 075502.

\bibitem{ABDELJAWAD2017528}
Abdeljawad, F., Lu, P., Argibay, N., Clark, B.~G., Boyce, B.~L. and Foiles,
  S.~M., {\em Acta Materialia\/}, 2017, {\bf 126}, 528 .

\bibitem{Mishin08a}
Mishin, Y., Boettinger, W.~J., Warren, J.~A. and McFadden, G.~B.,
  \enquote{Thermodynamics of grain boundary premelting in alloys. \mbox{I.
  Phase} field modeling,} Acta Mater. (2008), submitted as Part I of this work.

\bibitem{Frolov:2015ab}
Frolov, T. and Mishin, Y., {\em J. Chem. Phys.\/}, 2015, {\bf 143}, 044706.

\bibitem{Rickman201388}
Rickman, J., Chan, H., Harmer, M. and Luo, J., {\em Surface Science\/}, 2013,
  {\bf 618}, 88 .

\bibitem{Rickman20161}
Rickman, J., Harmer, M. and Chan, H., {\em Surface Science\/}, 2016, {\bf 651},
  1 .

\bibitem{Rickman2016225}
Rickman, J. and Luo, J., {\em Curr. Opin. Solid State Mater. Sci.\/}, 2016,
  {\bf 20}, 225 .

\bibitem{PhysRevB.90.144102}
Gao, Q. and Widom, M., {\em Phys. Rev. B\/}, Oct 2014, {\bf 90}, 144102.

\bibitem{Cahn82a}
Cahn, J.~W., {\em J. Physique Colloques\/}, 1982, {\bf 43}, 199.

\bibitem{Rottman1988a}
Rottman, C., {\em J. de Physique Colloque\/}, 1988, {\bf 49}, 313.

\bibitem{Hart:1972aa}
Hart, E.~W., in: {\em Nature and behavior of grain boundaries\/}, edited by
  H.~Hu, Plenum, New York, 1972  155--170.

\bibitem{Ratanaphan2015346}
Ratanaphan, S., Olmsted, D.~L., Bulatov, V.~V., Holm, E.~A., Rollett, A.~D. and
  Rohrer, G.~S., {\em Acta Materialia\/}, 2015, {\bf 88}, 346 .

\bibitem{Janssens06}
Janssens, K. G.~F., Olmsted, D., Holm, E.~A., Foiles, S.~M., Plimpton, S.~J.
  and Derlet, P.~M., {\em Nature Materials\/}, 2006, {\bf 5}, 124.

\bibitem{Olmsted07a}
Olmsted, D.~L., Foiles, S.~M. and Holm, E.~A., 2007, {\bf 57}, 1161.

\bibitem{Olmsted2011}
Olmsted, D.~L., Buta, D., Adland, A., Foiles, S.~M., Asta, M. and Karma, A.,
  {\em Phys. Rev. Lett.\/}, Jan 2011, {\bf 106}, 046101.

\bibitem{KIM20111152}
Kim, H.-K., Ko, W.-S., Lee, H.-J., Kim, S.~G. and Lee, B.-J., {\em Scripta
  Materialia\/}, 2011, {\bf 64}, 1152 .

\bibitem{MORITA19971053}
Morita, K. and Nakashima, H., {\em Materials Science and Engineering: A\/},
  1997, {\bf 234}, 1053 .

\bibitem{PhysRevB.85.064108}
Tschopp, M.~A., Solanki, K.~N., Gao, F., Sun, X., Khaleel, M.~A. and
  Horstemeyer, M.~F., {\em Phys. Rev. B\/}, Feb 2012, {\bf 85}, 064108.

\bibitem{doi:10.1080/13642818908211183}
Wolf, D., {\em Philosophical Magazine Part B\/}, 1989, {\bf 59}, 667.

\bibitem{doi:10.1080/01418619008244790}
Wolf, D., {\em Philosophical Magazine A\/}, 1990, {\bf 62}, 447.

\bibitem{doi:10.1063/1.347741}
Wolf, D., {\em Journal of Applied Physics\/}, 1991, {\bf 69}, 185.

\bibitem{PhysRevB.64.174101}
Ye\ifmmode~\mbox{\c{s}}\else \c{s}\fi{}illeten, D. and Arias, T.~A., {\em Phys.
  Rev. B\/}, Oct 2001, {\bf 64}, 174101.

\bibitem{HAHN2016108}
Hahn, E.~N., Fensin, S.~J., Germann, T.~C. and Meyers, M.~A., {\em Scripta
  Materialia\/}, 2016, {\bf 116}, 108 .

\bibitem{doi:10.1080/01418618308243118}
Tasker, P.~W. and Duffy, D.~M., {\em Philos. Mag. A\/}, 1983, {\bf 47}, L45.

\bibitem{Wolf82}
Wolf, D., {\em J. Phys. Colloques\/}, 1982, {\bf 43}, 45.

\bibitem{doi:10.1080/01418618208236206}
Sun, C.~P. and Balluffi, R.~W., {\em Philosophical Magazine A\/}, 1982, {\bf
  46}, 49.

\bibitem{doi:10.1080/01418618608242811}
Duffy, D.~M. and Tasker, P.~W., {\em Philos. Mag. A\/}, 1986, {\bf 53}, 113.

\bibitem{DUFFY84a}
Duffy, D.~M. and Tasker, P.~W., {\em J. Am. Ceram. Soc\/}, 1984, {\bf 67}, 176.

\bibitem{Phillpot1992}
Phillpot, S.~R. and Rickman, J.~M., {\em The Journal of Chemical Physics\/},
  1992, {\bf 97}, 2651.

\bibitem{Phillpot1994}
Phillpot, S.~R., {\em Phys. Rev. B\/}, Mar 1994, {\bf 49}, 7639.

\bibitem{Alfthan06}
\mbox{von Alfthan}, S., Haynes, P.~D., Kashi, K. and Sutton, A.~P., {\em Phys.
  Rev. Lett.\/}, 2006, {\bf 96}, 055505.

\bibitem{Alfthan07}
\mbox{von Alfthan}, S., Kaski, K. and Sutton, A.~P., {\em Phys. Rev. {\rm
  B}\/}, 2007, {\bf 76}, 245317.

\bibitem{PhysRevB.80.174102}
Zhang, J., Wang, C.-Z. and Ho, K.-M., {\em Phys. Rev. B\/}, Nov 2009, {\bf 80},
  174102.

\bibitem{Chua:2010uq}
Chua, A. L.~S., Benedek, N.~A., Chen, L., Finnis, M.~W. and Sutton, A.~P., {\em
  Nat Mater\/}, 05 2010, {\bf 9}, 418.

\bibitem{Frolov2013}
Frolov, T., Olmsted, D.~L., Asta, M. and Mishin, Y., {\em Nat. Commun.\/},
  2013, {\bf 4}, 1899.

\bibitem{Frolov2013PRL}
Frolov, T., Divinski, S.~V., Asta, M. and Mishin, Y., {\em Phys. Rev. Lett.\/},
  Jun 2013, {\bf 110}, 255502.

\bibitem{PhysRevB.92.020103}
Frolov, T., Asta, M. and Mishin, Y., {\em Phys. Rev. B\/}, Jul 2015, {\bf 92},
  020103.

\bibitem{Frolov2016}
Frolov, T., Asta, M. and Mishin, Y., {\em Curr. Opin. Solid State Mater.
  Sci.\/}, 2016, {\bf 20}, 308 .

\bibitem{Zhu2018}
Zhu, Q., Samanta, A., Li, B., Rudd, R.~E. and Frolov, T., {\em Nat. Commun.\/},
  2018, {\bf 9}, 467.

\bibitem{Demkowicz2015}
Yu, W. and Demkowicz, M., {\em Journal of Materials Science\/}, 2015, {\bf 50},
  4047.

\bibitem{Novoselov2016276}
Novoselov, I. and Yanilkin, A., {\em Computational Materials Science\/}, 2016,
  {\bf 112, Part A}, 276 .

\bibitem{Han2017}
Han, J., Vitek, V. and Srolovitz, D.~J., {\em Acta Materialia\/}, 2017, {\bf
  133}, 186 .

\bibitem{Frolov2018Nanoscale}
Frolov, T.~V., Setyawan, W., Kurtz, R., Marian, J., Oganov, A., Rudd, R.~E. and
  Zhu, Q., {\em Nanoscale\/}, 2018, ~--.

\bibitem{doi:10.1063/1.2210932}
Oganov, A.~R. and Glass, C.~W., {\em The Journal of Chemical Physics\/}, 2006,
  {\bf 124}, 244704.

\bibitem{Marinica2013}
Marinica, M.~C., Ventelon, L., Gilbert, M.~R., Proville, L., Dudarev, S.~L.,
  Marian, J., Bencteux, G. and Willaime, F., {\em Journal of Physics: Condensed
  Matter\/}, 2013, {\bf 25}, 395502.

\bibitem{Zhou2001}
Zhou, X., Wadley, H., Johnson, R., Larson, D., Tabat, N., Cerezo, A.,
  Petford-Long, A., Smith, G., Clifton, P., Martens, R. and Kelly, T., {\em
  Acta Materialia\/}, 2001, {\bf 49}, 4005 .

\bibitem{Kurtz2014}
Setyawan, W. and Kurtz, R.~J., {\em Journal of Physics: Condensed Matter\/},
  2014, {\bf 26}, 135004.

\bibitem{swissW2016}
Scheiber, D., Pippan, R., Puschnig, P. and Romaner, L., {\em Modelling and
  Simulation in Materials Science and Engineering\/}, 2016, {\bf 24}, 035013.

\bibitem{Setyawan2012558}
Setyawan, W. and Kurtz, R.~J., {\em Scripta Materialia\/}, 2012, {\bf 66}, 558
  .

\bibitem{CAMPBELL19993977}
Campbell, G., Belak, J. and Moriarty, J., {\em Acta Mater.\/}, 1999, {\bf 47},
  3977 .

\bibitem{Zhou-PRL-2014}
Zhou, X.-F., Dong, X., Oganov, A.~R., Zhu, Q., Tian, Y. and Wang, H.-T., {\em
  Phys. Rev. Lett.\/}, Feb 2014, {\bf 112}, 085502.

\bibitem{Zhu-PRB-2013}
Zhu, Q., Li, L., Oganov, A.~R. and Allen, P.~B., {\em Phys. Rev. B\/}, May
  2013, {\bf 87}, 195317.

\bibitem{Zhu-JCP-2014}
Zhu, Q., Sharma, V., Oganov, A.~R. and Ramprasad, R., {\em The Journal of
  Chemical Physics\/}, 2014, {\bf 141}, 154102.

\bibitem{Lyakhov-CPC-2013}
Lyakhov, A.~O., Oganov, A.~R., Stokes, H.~T. and Zhu, Q., {\em Computer Physics
  Communications\/}, 2013, {\bf 184}, 1172 .

\bibitem{Plimpton95}
Plimpton, S., {\em J. Comput. Phys.\/}, 1995, {\bf 117}, 1.

\bibitem{Zhu-PRB-2015}
Zhu, Q., Oganov, A.~R., Lyakhov, A.~O. and Yu, X., {\em Phys. Rev. B\/}, Jul
  2015, {\bf 92}, 024106.

\bibitem{PhysRevB.69.172102}
Lan\ifmmode~\mbox{\c{c}}\else \c{c}\fi{}on, F., Radetic, T. and Dahmen, U.,
  {\em Phys. Rev. B\/}, May 2004, {\bf 69}, 172102.

\bibitem{PhysRevLett.89.085502}
Radetic, T., Lan\ifmmode~\mbox{\c{c}}\else \c{c}\fi{}on, F. and Dahmen, U.,
  {\em Phys. Rev. Lett.\/}, Aug 2002, {\bf 89}, 085502.

\bibitem{Hirth96}
Hirth, J.~P. and Pond, R.~C., {\em Acta Mater.\/}, 1996, {\bf 44}, 4749.

\bibitem{Pond03a}
Pond, R.~C. and Celotto, S., {\em Int. Mater. Rev.\/}, 2003, {\bf 48}, 225.

\bibitem{HIRTH2013749}
Hirth, J., Pond, R., Hoagland, R., Liu, X.-Y. and Wang, J., {\em Progress in
  Materials Science\/}, 2013, {\bf 58}, 749 .

\bibitem{Larche_Cahn_78}
Larche, F.~C. and Cahn, J.~W., {\em Acta Metall.\/}, 1978, {\bf 26}, 1579.

\bibitem{doi:10.1063/1.448644}
Mullins, W.~W. and Sekerka, R.~F., {\em The Journal of Chemical Physics\/},
  1985, {\bf 82}, 5192.

\bibitem{Voorhees20041}
Voorhees, P. and Johnson, W.~C., volume~59 of {\em Solid State Physics\/},
  Academic Press, 2004  1 -- 201.

\bibitem{doi:10.1080/01418610208240038}
Campbell, G.~H., Kumar, M., King, W.~E., Belak, J., Moriarty, J.~A. and Foiles,
  S.~M., {\em Philosophical Magazine A\/}, 2002, {\bf 82}, 1573.

\bibitem{Medlin2017383}
Medlin, D., Hattar, K., Zimmerman, J., Abdeljawad, F. and Foiles, S., {\em Acta
  Materialia\/}, 2017, {\bf 124}, 383 .

\bibitem{doi:10.1063/1.4954066}
Abdeljawad, F., Medlin, D.~L., Zimmerman, J.~A., Hattar, K. and Foiles, S.~M.,
  {\em Journal of Applied Physics\/}, 2016, {\bf 119}, 235306.

\bibitem{PhysRevLett.120.085702}
Yang, S., Zhou, N., Zheng, H., Ong, S.~P. and Luo, J., {\em Phys. Rev.
  Lett.\/}, Feb 2018, {\bf 120}, 085702.

\bibitem{OBrien2018}
O'Brien, C.~J., Barr, C.~M., Price, P.~M., Hattar, K. and Foiles, S.~M., {\em
  Journal of Materials Science\/}, Feb 2018, {\bf 53}, 2911.

\bibitem{doi:10.1063/1.4880715}
Frolov, T., {\em Appl. Phys. Lett.\/}, 2014, {\bf 104}, 211905.

\bibitem{YIN2018141}
Yin, J., Wang, Y., Yan, X., Hou, H. and Wang, J.~T., {\em Computational
  Materials Science\/}, 2018, {\bf 148}, 141 .

\bibitem{ARAMFARD2018304}
Aramfard, M. and Deng, C., {\em Acta Materialia\/}, 2018, {\bf 146}, 304 .

\bibitem{doi:10.1021/ar800217x}
Erdemir, D., Lee, A.~Y. and Myerson, A.~S., {\em Accounts of Chemical
  Research\/}, 2009, {\bf 42}, 621, PMID: 19402623.

\bibitem{0965-0393-18-1-015012}
Stukowski, A., {\em Modell. Simul. Mater. Sci. Eng.\/}, 2010, {\bf 18}, 015012.

\end{thebibliography}
\end{document}